\magnification=1100

\hsize 17truecm
\vsize 23truecm

\font\twelvec=msbm10 at 10pt
\font\sevenc=msbm10 at 7pt
\font\fivec=msbm10 at 5pt

\newfam\co
\textfont\co=\twelvec
\scriptfont\co=\sevenc
\scriptscriptfont\co=\fivec

\def\ad{\mathop{\rm ad}\nolimits}

\def\Const{\mathop{\rm Const.}\nolimits}
\def\det{\mathop{\rm det}\nolimits}
\def\exp{\mathop{\rm exp}\nolimits}
\def\dist{\mathop{\rm dist}\nolimits}

\def\Id{\mathop{\rm Id}\nolimits}

\def\id{\mathop{\rm Id}\nolimits}

\def\Hom{\mathop{\rm Hom}\nolimits}

\def\lim{\mathop{\rm lim}\nolimits}

\def\Ran{\mathop{\rm Ran}\nolimits}

\def\sup{\mathop{\rm sup}\nolimits}
\def\inf{\mathop{\rm inf}\nolimits}
\def\sgn{\mathop{\rm sgn}\nolimits}
\def\SL{\mathop{\rm SL}\nolimits}

\def\neigh{\mathop{\rm neigh}\nolimits}

\def\WF{\mathop{\rm WF}\nolimits}

\def\Sum{\displaystyle\sum}
\def\e{\mathop{\rm \varepsilon}\nolimits}

\baselineskip 15pt
%\end
\centerline{\bf THE SEMICLASSICAL MAUPERTUIS-JACOBI CORRESPONDENCE}

\centerline{\bf FOR QUASI-PERIODIC HAMILTONIAN FLOWS:}

\centerline{\bf STABLE AND UNSTABLE SPECTRA}
\bigskip
\centerline{Sergey DOBROKHOTOV ${}^{(*)}$ {\it\&} Michel ROULEUX ${}^{(**)}$}
\bigskip
\centerline {${}^{(*)}$ Institute for Problems in Mechanics of Russian Academy of Sciences}

\centerline {Prosp. Vernadskogo 101-1, Moscow, 119526, Russia, dobr@ipmnet.ru} 

\centerline {${}^{(**)}$ Centre de Physique Th\'eorique and Universit\'e du Sud Toulon-Var, UMR 7332}

\centerline {Campus de Luminy, Case 907, 13288 Marseille Cedex 9, France, rouleux@univ-tln.fr}
\bigskip
\noindent {\bf Abstract}: We investigate semi-classical properties of
Maupertuis-Jacobi correspondence in 2-D for families of Hamiltonians $(H_\lambda(x,\xi), {\cal H}_\lambda(x,\xi))$, 
when ${\cal H}_\lambda(x,\xi)$ is
the perturbation of completely integrable Hamiltonian $\widetilde{\cal H}$ veriying some isoenergetic non-degeneracy conditions.
Assuming the Weyl $h$-PDO $H^w_\lambda$ has only discrete spectrum near $E$, and
the energy surface $\{\widetilde{\cal H}={\cal E}\}$ is separated by some pairwise disjoint lagrangian tori, we show that most of eigenvalues for 
$\widehat H_\lambda$ near $E$ are asymptotically degenerate as $h\to0$. 
This applies in particular for the determination of trapped modes by an island, in the linear theory of water-waves.
We also consider quasi-modes localized near rational tori. Finally, we discuss breaking of Maupertuis-Jacobi correspondence 
on the equator of Katok sphere.
\medskip
\noindent {\bf 0. Introduction}
\medskip
Let $M$ be a smooth manifold of dimension $d$, and ${\cal H},H\in C^\infty(T^*M)$ two Hamiltonians sharing a non critical
energy surface $\Sigma=\{{\cal H}={\cal E}\}=\{H=E\}$. Then ${\cal H},H$ have the same integral curves on $\Sigma$, up to a reparametrization 
of time $dt={\cal G}(\tau)d\tau$, where ${\cal G}$ depends on the initial condition. In other words, the Hamilton vector fields are related 
by $X_{\cal H}={\cal G}(\tau)X_H$; this theorem is due to Godbillon and its proof simplified by Weinstein [AbM].
Another simple proof follows from the fact that there is $c\in C^\infty(T^*M)$ elliptic such that
$${\cal H}(x,\xi)-{\cal E}=c(x,\xi;{\cal E},E)(H(x,\xi)-E)\leqno(0.1)$$
We say that the pair $({\cal H},H)$ satisfies Maupertuis-Jacobi correspondence (henceforth MJC) at energies $({\cal E},E)$. 
Of particular interest are the following examples:

(1) $|\xi|_g^2$ is a smooth Riemannian metric on $T^*M$, $V$ a smooth potential,
$$H(x,\xi)=|\xi|_g^2+V(x), \quad {\cal H}(x,\xi)={|\xi|_g^2\over E-V(x)}\leqno(0.2)$$
provided $E>\sup _M V(x)$, and ${\cal E}=1$. 
Parametrizations $t$ and $\tau$ are then related by $d\tau=(E-V)dt$; 
this was used by Levi-Civita in connexion with Kepler problem. 

(2) $M$ is diffeomorphic to an annulus of ${\bf R}^2$, 
endowed with a Liouville metric with a second quadratic integral ${\cal H}(x,\xi)=|\xi|^2g(x)$,
${\cal E}=1$, and 
$$H(x,\xi)=|\xi|(1+\mu(x) \xi^2)\tanh(D(x)|\xi|)\leqno(0.3)$$ 
is the dispersion relation governing gravity waves in linear
hydrodynamics, with depth $D(x)$ and surface tension $\mu(x)$, see [DoRo].

Maupertuis-Jacobi correspondence plays an important r\^ole,  
when the solutions $(\widetilde x(\tau),\widetilde\xi(\tau))$ of the Hamiltonian system
$(\dot{\widetilde x}(\tau),\dot{\widetilde\xi}(\tau))=X_{\cal H}(\widetilde x(\tau),\widetilde \xi(\tau))$
parametrizing a $d$-dimensional
torus $\Lambda\subset\Sigma$, are periodic, or quasi-periodic~: 
$(\widetilde x(\tau), \widetilde \xi(\tau))=(x^0(\widetilde\omega\tau+\varphi),\xi^0(\widetilde\omega\tau+\varphi))$.
Here $\widetilde\omega$ is a vector of periods and $(x^0(\varphi),\xi^0(\varphi))$ 
smooth functions on ${\bf T}^d$, ${\bf T}={\bf R}/2\pi{\bf Z}$. So $\Lambda$ is invariant under
both $X_{{\cal H}}$ and $X_{H}$. Assuming a Diophantine condition on $\widetilde\omega$, MJC induces on $\Lambda$
a quasi-periodic motion for $H$, with frequency vector $\omega=\widetilde\omega/\langle{\cal G}\rangle$ where 
$\langle{\cal G}\rangle$ denotes the average of ${\cal G}$ over $\Lambda$ [DoRo,Theorem 1.4]. Let $I_j=\oint_{\gamma_j}\xi dx$, $1\leq j\leq d$ 
be the action variables over a set of fundamental cycles $\gamma_j\subset\Lambda=\Lambda^I$, $I=(I_1,\cdots,I_d)$.

In [DoRo], we proved also the following~: Let 
$0<\delta<1$. Then in a $h^{\delta/2}$-neighborhood of $\Lambda$ in $T^*M$, there is 
a family $\Lambda^J$ of tori, labelled by their action variables $J=J_k(h)$ for all possible $k\in{\bf Z}^d$ satisfying $|kh-I|\leq h^\delta$, which
have the properties, that they verify Bohr-Sommerfeld-Maslov quantization condition, and are
quasi-invariant under $X_H$ with an accuracy ${\cal O}(h^\infty)$. These tori can be quantized, and thus give
raise to a spectral series near $E$ for ``any'' $h$-PDO $H^w$ with principal symbol $H$. 

Thus MJC transfers some knowledge relative to properties of the classical flow for Hamiltonian ${\cal H}(x,\xi)$, 
to properties of semi-classical spectrum for $H$.
The simplest way is to
think of ${\cal H}=\widetilde{\cal H}$ as being integrable near ${\cal E}$, 
i.e. being (locally) a function of some action variables $I\subset{\bf R}^d$ alone.

So let $(\widetilde{\cal H},\widetilde H)$ satisfy MJC at energies $({\cal E},E)$,
${\cal H}'\in C^\infty(T^*M)$, $\lambda$
be a small coupling constant, 
and ${\cal H}_\lambda=\widetilde{\cal H}+\lambda {\cal H}'$. 
It can happen that there corresponds a smooth
family of Hamiltonians $H_\lambda=\widetilde H(x,\xi)+\lambda H'(x,\xi;\lambda,{\cal E},E)$ such that 
$({\cal H}_\lambda,H_\lambda)$ satisfy MJC at energies $({\cal E},E)$ for small $\lambda$;
or conversely, given $H_\lambda=\widetilde H+\lambda H'$, that
${\cal H}_\lambda=\widetilde{\cal H}(x,\xi)+\lambda {\cal H}'(x,\xi;\lambda,{\cal E},E)$ and $H_\lambda$
satisfy MJC at $({\cal E},E)$. 

This is the case for (0.2); this holds also in case of the dispersion relation (0.3), $\widetilde{\cal H}(x,\xi)=|\xi|^2g(x)$
a Liouville metric with a second quadratic integral, and
${\cal H}_\lambda=|\xi|^2g(x)+\lambda|\xi|^2g'(x)$,
provided the depth profile $D=D(x;E,\lambda)$  is conveniently
chosen as a function of the metric $g(x)$, or vice-versa; see [DoRo,Proposition 4.1] and its proof.

Assume the isoenergetic non degeneracy condition on $\widetilde\omega(I)={\partial\widetilde{\cal H}\over\partial I}$ holds in $\Sigma$, i.e. 
$$\det \pmatrix{{\partial\widetilde\omega(I)\over\partial I}&\widetilde\omega(I)\cr {}^t\widetilde\omega(I)&0\cr}\neq0\leqno(0.5)$$
which means that $I\mapsto[\widetilde\omega(I)]$ 
restricted to the energy surface $\widetilde{\cal H}={\cal E}$ is a (local) isomorphism on the projective space.
When $\sigma>d-1$, and for $c>0$ small enough, we define a KAM set on $\Sigma$, as the Cantor set~:
$$K_{c,\sigma}=\{I\in\Sigma=\widetilde{\cal H}^{-1}({\cal E}): \forall k\in{\bf Z}^d\setminus0, \ |\langle k,\widetilde\omega(I)\rangle|
\geq c|k|^{-\sigma}\}\leqno(0.6)$$
whose complement has a small measure (of order $c$) as $c\to0$.
For $I\in K_{c,\sigma}$, we know that the KAM torus $\Lambda(I)$ survives small perturbations ${\cal H}_\lambda=\widetilde{\cal H}+\lambda{\cal H}'$ 
of $\widetilde{\cal H}$, and
the Hamiltonian flow for ${\cal H}_\lambda$ is again quasi-periodic on a deformation
$\Lambda_\lambda(I)$ of $\Lambda(I)$ with a frequency vector proportional to $\widetilde\omega(I)$ 
(see [Bo,Theorem 1.2.2]).

Consider now the semi-classical case, and
let $H(x,\xi,h)=H_\lambda(x,\xi,h)$ belong to the usual class 
$$S^0(M)=\{H\in C^\infty(T^*M): |\partial_x^\alpha\partial_\xi^\beta H(x,\xi;h)|\leq C_{\alpha,\beta}\}$$
with asymptotics $H(x,\xi,h)\sim H_0(x,\xi)+hH_1(x,\xi)+h^2H_2(x,\xi)+\ldots$, as $h\to0$, and 
$$H^w(x,hD_x;h)u(x;h)=\int\int e^{i(x-y)\xi/h}H({x+y\over2},\xi;h)u(y)dyd\xi\leqno(0.7)$$ 
Function $H_0$ (or principal symbol of $H^w(x,hD_x;h)$,~) is the classical Hamiltonian, and $H_1$ the sub-principal symbol.
This makes always sense if $M={\bf R}^d$ or ${\bf T}^d$,
otherwise $H^w(x,hD_x;h)$ may be defined up to its principal and sub-principal symbols only (see [H\"o,Chap.XVIII] for 
a general discussion on PDO's on a manifold.~)

Using Theorem 1.1, we can construct a quasi-mode for $H^w_\lambda(x,hD_x;h)$ of infinite order, corresponding to 
asymptotic eigenvalues $E_k(h)$, for all lattice points $kh$ (possibly shifted by Maslov index), within a distance of any KAM set on $\Sigma$,
not exceeding $h^\delta$. The dimension of the span of the corresponding asymptotic eigenfunctions $\varphi_k(h)$ is about
$(2\pi h)^{-d}|K_{c,\sigma}|$, see Theorem 1.2.
This set of lattice points has several ``connected components'', separated by so-called ``resonance'' zones.
Following [CdV2], we call this part of the spectrum of the family
$H^w_\lambda$ the {\it stable spectrum} induced by Maupertuis-Jacobi correspondence.

In this paper we shall focus on the {\it unstable spectrum} of $H^w_\lambda$ instead, associated
with quasi-modes concentrated on rational Lagrangian tori, or elliptic periodic orbits, or with 
so-called Shnirelman quasi-modes concentrated on connected components of $\Sigma$ between KAM tori.

The latter play a particular important r\^ole when ${\cal H}$ is invariant under the involution $\Gamma:(x,\xi)\mapsto(x,-\xi)$, and 
some KAM tori $\Lambda$ and $\Gamma(\Lambda)$ are pairwise disjoint. 

Let us indeed formulate our first main result concerning quasi-modes supported between KAM tori. 
Given some $J(h)\subset{\bf N}$, with $|J(h)|\to\infty$ as $h\to0$, and
$J'(h)\subset J(h)$, we say that $J'(h)$ is of relative density 1 iff 
$\lim_{h\to0}{|J'(h)|\over|J(h)|}=1$.
We have~:
\medskip
\noindent{\bf Theorem 0.1}: {\it Assume $d=2$. Let $H\in S^0(M)$ be as above, and assume $H^w$  has only discrete spectrum in 
$I(h)=[E-h^\delta,E+h^\delta]$, $0<\delta<1$. 
Let $J(h)=\{j\in {\bf N}: \lambda_j(h)\in I(h)\}$ label the eigenvalues in $I(h)$, counted with multiplicity. 
On the other hand, let ${\cal H}$ be completely integrable, or a small perturbation of an integrable Hamiltonian $\widetilde{\cal H}$,
${\cal H}=\widetilde{\cal H}+\lambda{\cal H}'$, satisfying (0.5) so that KAM theory applies. 

Assume that $({\cal H},H)$ satisfy MJC at energies $({\cal E}, E)$, and 
$$\Sigma=\{H=E\}=\{{\cal H}={\cal E}\} \ \hbox{is compact, diffeomorphic to} \  
{\bf T}^2\times{\bf S}^1 \ \hbox{or} \  {\bf T}^2\times[0,1]\leqno(0.8)$$
so that it is separated by any 2 pairs of invariant tori.  

Assume also ${\cal H}$ is time-reversal invariant, i.e. invariant under the involution 
$\Gamma:(x,\xi)\mapsto(x,-\xi)$, and there exists 4 invariant tori (separated in phase-space) with diophantine frequency vectors
$$\Lambda_1, \ \Lambda_2=\Gamma(\Lambda_1), \ \Lambda_3, \ \Lambda_4=\Gamma(\Lambda_3)\leqno(0.9)$$
Then there exists $J'(h)\subset J(h)$ of relative density 1 such that 
$\forall j\in J(h): |\lambda_{j\pm1}(h)-\lambda_j(h)|={\cal O}(h^\infty)$.}
\smallskip
Thus, in a $h^\delta$-neighborhood of $E$, we can find 
a subsequence of ``density 1'' of asymptotically degenerate (at least double) eigenvalues of $H(x,hD_x,h)$, mod ${\cal O}(h^\infty)$.
This is the case for the Liouville metric above,
for which momentum tunneling was computed in [DoSh] using complex cycles
and shown to be exponentially small.
It would be of course much harder to understand tunneling properties for $H$ in a direct way.

We consider next the situation where
the frequency vector associated with flow of $X_{\cal H}$ on $\Lambda$ is rational (then we say for short that $X_{\cal H}$ has rational flow).
It may happen that the flow of $X_H$ on $\Lambda$ induced by MJC is again rational~; in this case we proved in [DoRo] that
it is possible to construct quasi-modes for $H^w$ in a $h^{1/2}$-nghbd of $\Lambda$. 
In general however, the flow of $X_H$ on $\Lambda$ is again conjugated to a linear flow, 
but with a frequency that generally depends on the initial condition. 
There is no canonical way to determine the motion on nearby 
tori $\Lambda^J$, but 
under some ellipticity condition, this motion can be identified with a Larmor precession in a varying magnetic field,
which allows to construct various types of quasi-modes, according to some components of a Reeb graph.

At last, we provide an example of breaking of semi-classical MJC, 
related to projectively equivalent Finsler symbols
on the sphere; this leads to spectral series for an operator of Aharonov-Bohm type.

These results were announced in [DoRo2]. 
\medskip
\noindent {\bf Acknowledgements}: We thank V.Ivrii for useful information about trace formulas, 
Ch.Duval and M.Taylor for an introduction to Katok sphere. The second author also wants to thank M.
Tsfasman for his kind hospitality in Laboratoire Poncelet, UMI 2615 CNRS, Moscow, where part of this work was done.
\bigskip
\noindent {\bf 1) Maupertuis-Jacobi correspondence and Shnirelman quasimodes}
\smallskip
We present here an extension to the semiclassical case of a construction by Shnirelman [Sh].
\smallskip
\noindent{\it a) Birkhoff normal form, and the semi-classical quantization near a Diophantine torus}.
\smallskip
Consider an Hamiltonian $H(x,\xi)\in C^\infty(T^*M)$, with $M=M^d$ a smooth manifold. 
Let $\Lambda$ be a smooth Lagrangian torus invariant under the Hamiltonian flow of $H$, conjugated
to a linear (Kronecker) flow on ${\bf T}^d$, with 
frequency vector $\omega$, $H|_{\Lambda}=E$. Assume also $i:\Lambda\to T^*M$ is a Lagrangian embedding, 
so that $H^1(\Lambda;{\bf R})$ is stable under small perturbations. 
Let $(\gamma_j)_{1\leq j\leq d}$ be basic cycles on
$\Lambda$. They determine action variables $I_j=\oint_{\gamma_j} pdx$, and also
Maslov indices  $\alpha_j$.
Introducing suitable action-angle coordinates as in [BeDoMa], [DoRo] (as a particular case of Darboux-Weinstein theorem), we get a 
canonical transformation
$$\widetilde\kappa:\neigh(\Lambda;T^*M)\to\neigh(\iota=0;T^*{\bf T}^d)\leqno(1.1)$$
which maps $\Lambda$ to the zero section in $T^*{\bf T}^d$, and such that when expressed in the coordinates $(\varphi,\iota)$
$$H=H|_\Lambda+\langle\omega,\iota\rangle+{\cal O}(|\iota|^2)\leqno(1.2)$$
When $\omega$ is Diophantine, this can be improved by BNF, so that 
in a new set of action-angle variables $(\varphi',\iota')$, which we construct 
by applying successively the averaging method,
$H$ becomes independent of $\varphi'$ up to ${\cal O}(|\iota'|^{N+1})$.
More precisely for each $N=1,2,3,\cdots$, 
there is a smooth canonical map $\kappa_N:(\varphi',\iota')\mapsto (\varphi,\iota)$, $d\kappa_N|_{\iota=0}=\Id$, defined for  
$(\varphi',\iota')\in{\bf T}^d\times\neigh(\iota=0;{\bf R}^d)$, and
a polynomial $H_N(\iota')$ of degree $N$,
with $H_N(\iota')=E+\langle\omega,\iota'\rangle+{\cal O}(|\iota'|^2)$, $E=H|_\Lambda$, such that 
$$H(x,\xi)=H\circ\widetilde\kappa\circ\kappa_N(\varphi',\iota')=H_N(\iota')+{\cal O}(|\iota'|^{N+1})\leqno(1.3)$$
(with the convention $\kappa_N=\Id$ when $N=1$.~) See e.g. [DoRo,Thm 2.2]. 
The sequence of $\kappa_N(\varphi,\iota)$ is {\it nested}, in the sense that 
for all $N$, $\kappa_{N+1}(\varphi,\iota)-\kappa_N(\varphi,\iota)={\cal O}(|\iota|^{N+1})$. 
Given a sequence of nested $\kappa_N$ we can construct, by Borel procedure, a
canonical transformation $\kappa$ such that for all $N$, 
$\|\kappa(\varphi,\iota)-\kappa_N(\varphi,\iota)\|={\cal O}(|\iota|^{N+1})$. 

Denote by $\Lambda(J',N)$ the preimage of 
${\bf T}^d\times\{J'=I+\iota'\}$ by $\widetilde\kappa\circ\kappa_N$. Note that when $\iota'\neq0$, $\Lambda(J',N)$ is only quasi-invariant 
with respect to $X_H$. Moreover, the classical action $\oint_{\gamma_j}p\, dx$ over a fundamental cycle $\gamma_j$ on $\Lambda(J',N)$
is related with $I^0=\oint_{\gamma^0_j}p\, dx$ over a fundamental cycle $\gamma^0_j$ on $\Lambda_0$ by 
$\oint_{\gamma_j}p\, dx= I^0+\iota'+{\cal O}(\iota'^2)$ (see [DoRo,Corollary 2.4]). 

Consider now the usual quantization $H^w(x,hD_x)$ of $H(x,\xi)$ as in (0.5).
In Appendix A.a we review some concepts of microlocal analysis; ``anisotropic admissible boxes'' 
$\Pi_{\rho^0}^\delta\subset T^*{\bf R}^d\times T^*M$, 
centered at some $\rho^0=(\varphi^0,I^0,x^0,\xi^0)\in T^*{\bf R}^d\times T^*M$, are of the form
$$\Pi_{\rho^0}^\delta=\{(\varphi,I,x,\xi): |\varphi_j-\varphi_j^0|\leq c, |I_j-I_j^0|\leq ch^\delta,|x_j-x_j^0|\leq ch^{\delta/2}, 
\ |\xi_j-\xi_j^0|\leq ch^{\delta/2}\}, \ c>0$$
Let $K\in{\cal I}^m({\bf R}^d\times M)$ be a Lagrangian distribution; if there exists 
an admissible box $\Pi_{\rho^0}^\delta$ around $\rho^0$ such that $K$ is ``negligible'' in $\Pi_{\rho^0}^\delta$, 
we write $\rho^0\notin{\WF'}^\delta K$, which defines a closed subset of $T^*{\bf R}^d\times T^*M$ called the (anisotropic)
{\it wave-front set} or {\it oscillation front}. In App.A.c we prove the (probably well known):
\medskip
\noindent {\bf Theorem 1.1}:   
{\it Let $\Lambda=\Lambda_0$ be a Lagrangian torus with Diophantine frequencies 
with the actions
$I^0_j=\oint_{\gamma^j}p\, dx$ along a set of fundamental cycles, and Maslov indices $\alpha$. Let also $H(x,\xi;h)=H_0(x,\xi)+h^2H_1(x,\xi)+\cdots$ 
be a classical symbol on $T^*M$ microlocally defined near $\Lambda_0$, with zero sub-principal symbol $H_1$, 
and assume the Hamilton vector field $X_{H_0}$ is tangent to $\Lambda_0\subset H_0^{-1}(E)$. Then
we can find a system of action-angle coordinates near $\Lambda_0$, such that there are
${\rm (i)}$ a (microlocally) unitary FIO operator $U^h:C_0^\infty({\bf R}^d)\to C^\infty(M)$, with 
${\WF'}^\delta U^h\subset({\bf R}^d\times \{I^0\})\times \Lambda_0$, and ${\rm (ii)}$ 
a $h$-PDO $P(hD_{\varphi'};h)$ (whose full symbol depends only on action variables, modulo ${\cal O}(h^\infty)$ remainder terms), such that
$$H(x,hD_x;h)U^h-U^hP(hD_{\varphi'};h)\in{\cal I}^{-\infty}\bigl({\bf R}^d\times M\bigr)$$
where we have identified ${\WF'}^\delta U^h$ with the wave-front set of its Schwartz kernel.}
\smallskip
We stress that $U^h$ does not act upon semiclassical distributions defined on ${\bf T}^d$, 
but rather on those which are defined on a $h$-dependent number of sheets of its covering ${\bf R}^d$.
We call it the {\it generalized semi-classical BNF}, which extends previous results by
[We] and [CdV1], in case of the geodesic flow on a Riemannian manifold (see also [Sh], [Po], [HiSjVu]).
Note that the usual Maslov canonical operator, the main tool for constructing quasi-modes (see [Laz], or [DoRo] for a simpler proof)
has domain the set of semiclassical distributions microlocalized on tori which satisfy Bohr-Sommerfeld-Maslov quantization condition.
However, we show in App.A.c that Theorem 1.1 also implies existence of such quasi-modes, with quasi-energies 
$P(kh-I^0-h\alpha/4;h)$, $|k|h\leq ch^\delta$. To construct quasi-modes, one usually contents to ``freeze'' the action-variable at 
$\iota=0$. 
\medskip
\noindent{\it b) KAM sets and the semi-classical 2-D Maupertuis-Jacobi correspondence}.
\smallskip
Our first result, in the spirit of [CdV], is about the ``mass'' of the QM we can construct for $H_\lambda$ near energy $E$, 
when $({\cal H}_\lambda,H_\lambda)$
satisfy MJC at energies $({\cal E},E)$, and ${\cal H}_\lambda$ is the perturbation of a completely integrable semi-classical system.

Consider first the Hamiltonian ${\cal H}_\lambda$. As $X_{\widetilde{\cal H}}$ is completely integrable near ${\cal E}$, Arnold-Liouville-Mineur theorem
shows that in a system of action-angle coordinates $(\widetilde\varphi,\widetilde\iota)$, we have
$\widetilde{\cal H}\circ\kappa^{-1}(\widetilde\varphi,\widetilde\iota)=\widetilde{\cal H}(\widetilde\iota)$. 
Composing ${\cal H}_\lambda(x,\xi)=\widetilde{\cal H}(x,\xi)+\lambda {\cal H}'(x,\xi;\lambda)$ with $\kappa^{-1}$ 
gives a function ${\cal H}_\lambda(\widetilde\varphi,\widetilde\iota)=\widetilde{\cal H}(\widetilde\iota)+\lambda 
{\cal H}'(\widetilde\varphi,\widetilde\iota;\lambda)$. 

Let $\Lambda_0\subset\widetilde\Sigma=\{\widetilde{\cal H}={\cal E}\}$ 
be a Lagrangian integral manifold,
with frequency vector $\widetilde\omega_0=(\widetilde\omega_1^0,\widetilde\omega_2^0)$.
Changing $\widetilde\iota$ by a constant, we will assume that $\Lambda_0$ is given by $\widetilde\iota=0$. 
An application of the implicit function theorem shows that a neighborhood of $\Lambda_0$ in the
energy surface $\widetilde\Sigma=\widetilde{\cal H}^{-1}({\cal E})$ can be parametrized by $\widetilde\iota_2=\widetilde f(\widetilde\iota_1)$, 
where $\widetilde f$ is smooth near 0, $\widetilde f(0)=0$, $\widetilde f'(0)=-\widetilde\omega_1/\widetilde\omega_2$. 
Condition (0.5) takes the form $\widetilde f''(0)\neq0$.
This implies that for small $\mu$, the $X_{\widetilde{\cal H}}$-invariant tori 
$\Lambda_\mu=\{(\widetilde\iota_1,\widetilde\iota_2)=(\mu,f(\mu))\}\subset\widetilde{\cal H}^{-1}({\cal E})$ can be parametrized by the
corresponding rotation numbers $f'(\mu)$.
For fixed $\sigma>1$, and small $c>0$, define as in (0.6) the KAM set
$$K_{c,\sigma}=\{\mu\in\neigh(0): \ |f'(\mu)-{p\over q}|\geq{c\over q^{\sigma}}, \ \forall p\in{\bf Z}, \ 
\forall q\in{\bf N}\setminus0\}\leqno(1.9)$$
By the isoenergetic KAM theorem (see e.g. [Bo], [ArKoNe], [HiSjVu,Thm.7.5] for more precise statements),
there exists $C>0$ sufficiently large, and a smooth family of smooth maps
$\Psi_\lambda:{\bf T}^2\times\neigh(0)\to\neigh(\widetilde\iota=0;T^*{\bf T}^2)$, with
$|\lambda|\leq c^2/C$, such that for all $\mu\in K_{c,\sigma}$, the set 
$$\Lambda_{\mu,\lambda}=\{\Psi_\lambda(\widetilde\varphi,\mu): \widetilde\varphi\in {\bf T}^2\}\subset T^*{\bf T}^2\leqno(1.10)$$ 
is a Lagrangian torus which can be embedded in the energy surface $\Sigma_\lambda={\cal H}_\lambda^{-1}({\cal E})$ 
as a Lagrangian manifold close to $\Lambda_\mu$. 
The Hamilton flow
on $\Lambda_{\mu,\lambda}$ is conjugated to a linear flow with rotation number $f'(\mu)$, and frequency vector $\widetilde\omega_\mu$. Moreover 
Liouville measure of the complement in ${\cal H}_\lambda^{-1}({\cal E})$ of 
$\bigcup_{\mu\in K_{c,\sigma}}\Lambda_{\mu,\lambda}$, 
is ${\cal O}(c)$, uniformly for $|\lambda|\leq c^2/C$. 
Further we can arrange so that ${\cal H}_\lambda^{-1}({\cal E})$ is (locally) foliated by the $\Lambda_{\mu,\lambda}$,
for ${\mu\in\neigh(0)}$, although, when $\mu\notin K_{c,\sigma}$, $\Lambda_{\mu,\lambda}$ need not be invariant under $X_{{\cal H}_\lambda}$.

We turn now to the quantum case, and consider the $h$-PDO ${\cal H}^w(x,hD_x)$ on $L^2(M)$.
Let $\widetilde{\bf H}$
be the span of quasi-modes of ${\cal H}$ in ${\cal H}^{-1}([{\cal E}-h^\delta,{\cal E}+h^\delta])$, where we recall that 
${\cal H}^{-1}({\cal E})$ is compact and non singular. Assume also, for simplicity, that ${\cal H}^w(x,hD_x)$
is completely integrable near ${\cal E}$, i.e. there are 
${\cal H}_1^w(x,hD_x)$, \dots, ${\cal H}_{d-1}^w(x,hD_x)$
commuting 
with ${\cal H}(x,hD_x)$. Then the number of eigenvalues of ${\cal H}^w(x,hD_x)$, i.e. the joint spectrum of
these operators in $[{\cal E}-h^\delta,{\cal E}+h^\delta]$
is about $Ch^{\delta-d}$. Let $K_{c,\sigma}$ be a KAM set as in (1.9), and $K_1=K^1_{c,\sigma}$ the closure of the set of points of density 1 
in $K_{c,\sigma}$, so that $|K_{c,\sigma}\setminus K^1_{c,\sigma}|=0$. Let also $K'_1\subset{\bf R}^d$ be the image of this set by the 
inverse of the map $I\mapsto\omega(I)$. 
For small $\lambda$, consider now the quasi-integrable Hamiltonian ${\cal H}_\lambda$;
as is recalled in the discussion after Theorem 1.1 we can construct a family of quasi-modes for ${\cal H}_\lambda(x,hD_x)$ with 
quasi-energies in 
$[{\cal E}-h^\delta,{\cal E}+h^\delta]$, $|kh|\leq h^\delta/C$. We know [CdV2] 
that the span $\widetilde{\bf H}^1_\lambda$ of such quasi-modes satisfying $\dist (kh+\alpha h/4, K'_1)\leq h^\delta/C_1$ 
has dimension $\dim \widetilde{\bf H}^1_\lambda\sim h^{\delta-d}|K_1|$. 
This is the semi-classical analogue of the KAM set, and this part of the spectrum of ${\cal H}_\lambda(x,hD_x)$ is called the 
{\it stable} spectrum. 

Let now $({\cal H}_\lambda,H_\lambda)$ satisfy MJC at energies $({\cal E},E)$.
Recall from [DoRo] that, because of the Diophantine condition,
the flow of $H_\lambda$ on $\Lambda_{\mu,\lambda}$ is again conjugated to a linear flow, with vector of frequencies
$\omega=(\omega_1,\omega_2)$ proportional to the vector of frequencies 
$\widetilde\omega_{\mu}$ for the corresponding flow of ${\cal H}_\lambda$. 

Since everything depends smoothly on $\lambda$, 
without loss of generality, we can think 
below of $\lambda=0$,  and also $\mu=0$ for local constructions. So we shall omit the subscripts 
${\mu,\lambda}$ when unnecessary, and so to stress that this Hamiltonian stands for the principal symbol of a $h$-PDO, we
denote sometimes $H_\lambda$ by $H_0$. 

Again, with the help of Theorem 1.1 we can construct a family of quasi-modes $f_k$ for $H_\lambda(x,hD_x)$ with 
quasi-energies in $[E-h^\delta,E+h^\delta]$, $|kh|\leq h^\delta/C$, 
Because of (0.1), with any $h^\delta$-nghbhd of $\widetilde\Sigma$ corresponds a $h^\delta$-nghbhd of $\Sigma_\lambda$ of same size.
So we get easily:
\medskip
\noindent {\bf Theorem 1.2}: {\it Assume (1.9) and $({\cal H}_\lambda,H_\lambda)$ satisfy MJC at energies $({\cal E},E)$. Then 
there is a family of quasi-modes for $H_\lambda(x,hD_x)$ with 
quasi-energies in 
$[E-h^\delta,E+h^\delta]$, $|kh|\leq h^\delta/C$, and the span of such quasi-modes satisfying $\dist (kh+\alpha h/4, K'_1)\leq h^\delta/C_1$,
with $K'_1\subset{\bf R}^d$ as above, 
has dimension $\dim \widetilde{\bf H}^1_\lambda\sim h^{\delta-d}|K_1|$. }
\medskip
\noindent{\it c) A quasi-projector for 2-D Maupertuis-Jacobi correspondence}.
\smallskip
We are to construct a $h$-PDO which we call, according to Shnirelman, a ``quasi-projector'' 
for $H$ (though it has no reason to satisfy everywhere the relation $Q^2=Q$)
associated with the decomposition of $\Sigma_\lambda$ in 2 connected components. 

Actually, Shnirelman's construction was devised for a small perturbation of an integrable system, 
but as we show below, it extends readily to our setting, with semi-classical limit instead of high energy asymptotics. 
To this respect, the semi-classical limit
turns out to be easier, since we replace the scale of finite regularity in Sobolev spaces by an ordering in powers of $h$.

So let $\Lambda_i=\Lambda_{\mu_i}\subset\widetilde{\cal H}^{-1}({\cal E}), i=1,2$, be any pair of invariant 
tori with rotation numbers $f_i(\mu_i)$, with $f_i=f$ as above and
$\mu_i\in K_{c,\sigma}$, dividing $\widetilde\Sigma$ into 2 domains, that we will denote by $\widetilde\Sigma_3$ and $\widetilde\Sigma_4$
(for simplicity, we assume that $\widetilde\Sigma=\{\widetilde{\cal H}={\cal E}\}$ is diffeomorphic to ${\bf T}^2\times{\bf S}^1$, the case
${\bf T}^2\times[0,1]$ being similar).
Considering instead the sets $\Lambda_{\mu_i,\lambda}$ as in (1.10) for small $\lambda$, we obtain a corresponding partition 
$$\Sigma_\lambda=\Lambda_{\mu_1,\lambda}\cup\Sigma_{3,\lambda}\cup\Lambda_{\mu_2,\lambda}\cup\Sigma_{4,\lambda}\leqno(1.13)$$
with $\Sigma_{i,\lambda}$ open. 
Since $\Lambda_{\mu,\lambda}=\{\widetilde\iota_1=\mu, \widetilde\iota_2=\widetilde f(\widetilde\iota_1)\}$, 
we can assume that, locally near $\Lambda_{\mu_1,\lambda}$
$$\Sigma_{3,\lambda}=\{\widetilde\iota_1>\mu_1, \widetilde\iota_2=\widetilde f(\widetilde\iota_1)\}, \quad 
\Sigma_{4,\lambda}=\{\widetilde\iota_1<\mu_1, \widetilde\iota_2=\widetilde f(\widetilde\iota_1)\}\leqno(1.14)$$ 
and similarly near $\Lambda_{\mu_2,\lambda}$. We say that the part 
$\Sigma_{3,\lambda}$ of $\Sigma_\lambda$ 
belongs, locally, to the ``right hand side'' of $\Lambda=\Lambda_{\mu_1,\lambda}$, and $\Sigma_{4,\lambda}
\subset\{\widetilde\iota_1<\mu\}$ to its ``left hand side''. 

Assuming MJC $\Sigma_\lambda={\cal H}_\lambda^{-1}({\cal E})=H_\lambda^{-1}(E)$,
our purpose is to construct a quasi-projector for $H_\lambda$, from the classical dynamics of ${\cal H}_\lambda$, which in turn is determined by 
this of $\widetilde{\cal H}$. 
Due to the fact that MJC preserves Hamiltonian curves
(up to reparametrization of time), partition (1.13) is again invariant by the Hamiltonian flow of $H_\lambda$. 
Moreover we know that the action-angle coordinates $(\varphi,\iota)$ constructed near $\Lambda_{\mu_i,\lambda}$
as in (1.1), have the property that $\iota-\widetilde\iota={\cal O}(\widetilde\iota^2)$ (see [BeDoMa]).
It follows easily that the defining functions for $\Sigma_{j,\lambda}$ in coordinates $(\varphi,\iota)$, are again of the form (1.14), when $\widetilde f$
replaced with another smooth $f$. The surfaces ${\cal I}_1$ near $\Lambda_1$, 
(resp. ${\cal I}_2$ near $\Lambda_2$), given by $\{\iota_1=0\}$ in the local action-angle coordinates above,
are transverse to $\Sigma$, $\Sigma$ intersects $F_j$ along $\Sigma_j$, and ${\cal I}_j$ along $\Lambda_j$.
So let ${\cal V}$ be an open neighborhood of $\Sigma$ in $T^*M$, separated by ${\cal I}_1$ and ${\cal I}_2$, and $F_3,F_4$ its connected
components. 

Recall from Definition a.2 that, if $A$ denotes an admissible $h$-PDO with symbol $a\in S^0_\delta(M)$,
we say $a\in S^0(F^{\delta/2})$, iff for all $\rho\notin F$, $\widehat A$ is negligible in a $h^{\delta/2}$ nghbhd of $\rho$. 

First we construct the quasi-projector locally on $F_3$, ``the right hand side'' of $\Lambda$, in the class of symbols 
$\widetilde S^m_{\delta}({\bf T}^d)$ defined in (A.5). 
We can think of $\Lambda$ as the zero-section of 
$T^*{\bf T}^d$, and  denote $a\in \widetilde S^0(F^{\delta})$ instead of $a\in S^0(F^{\delta/2})$ to emphasize the use of 
action-angle coordinates. 
\medskip
\noindent {\bf Lemma 1.3}: {\it There exists a $h$-PDO $Q_\Lambda=Q_\Lambda(\varphi,hD_\varphi,h)$
whose symbol $q_\Lambda$ belongs to $\widetilde S^0_{\delta}({\bf T}^2)$, $0<\delta<1$ and such that~:

\noindent (i) $Q_\Lambda$ ``almost commutes'' with $H$, i.e. for all $N$ there is $C_{N}>0$ such that in local operator norm,
for $h>0$ small enough~:
$$\|{i\over h}[Q_\Lambda,H]\|\leq C_{N}h^N \leqno(1.15)$$
(ii) Consider $a,b\in\widetilde S^0_{\delta}({\bf T}^2)$ be supported outside a sufficiently large $h^{\delta/2}$-nghbd of $\{\iota_1=0\}$,
$a\in \widetilde S^0(F_3^{\delta})$, $b\in \widetilde S^0(F_4^{\delta})$. Let 
$A,B\in\widetilde L^0_{\delta}({\bf T}^d)$ the corresponding operators.
Then for all $N>0$, there exists $C_N>0$ such that~:
$$\|A(\Id-Q_\Lambda)\|\leq C_{N}h^N, \quad  \|BQ_\Lambda\|\leq C_{N}h^N\leqno(1.16)$$
(iii) If $a\in S^0_0({\bf T}^d)$, and $A\in L^0_0({\bf T}^d)$ is the corresponding operator, we have 
$$\|{i\over h}[A,Q_\Lambda]\|\leq Ch^{-\delta}\leqno(1.17)$$
The same holds for $Q_\Lambda^*$. }
\smallskip
\noindent {\it Sketch of proof}: For simplicity we shall identify a $h$-PDO with its symbol, denoting them by the same letter.
Let $\chi(\iota)$ be a smooth cutoff, equal to 1 in a small but fixed
neighbd of 0. From (1.1) it follows that $\chi$ is again a cut-off in $T^*M$ equal to 1 in a nghbhd of $\Sigma$, locally near $\Lambda$. 
Let also $\Phi\in C^\infty({\bf R})$, $\Phi'(\eta_1)\geq0$, $\Phi(\eta_1)=0$ for $\eta_1<-1$, $\Phi(\eta_1)=1$ for $\eta_1>1$,
and set
$$q_\Lambda(\varphi,\iota,h)=\chi(\iota)\Phi({\iota_1\over h^{\delta}})\leqno(1.18)$$
We have $q_\Lambda\in \widetilde S^0_{\delta}({\bf T}^d)$, 
$1-q_\Lambda\in \widetilde S^0_{\delta}({\bf T}^d)$, and by Proposition A.3, $q_\Lambda\in \widetilde S^0_{\delta}(F^\delta), F=\{\iota_1>0\}$ and
$1-q_\Lambda\in \widetilde S^0_{\delta}(\check F^\delta), \check F=\{\iota_1<0\}$. Here $F=F_3,\check F=F_4$ represent 
locally the splitting of ${\cal V}$ by $\Lambda$.
If $H$ would depend on action variables $\iota$ only, the corresponding operator $q_\Lambda(\varphi,hD_\varphi;h)$ would satisfy (i) since
$[q_{\Lambda},H]=0$. 
But applying Theorem 1.1 to the Diophantine torus $\Lambda$,
we can find a $h$-PDO $Q_\Lambda=U^hq_\Lambda (U^h)^{-1}$ microlocally in a $h^{\delta/2}$ nghbhd of $\Lambda$, which verifies 
$$[Q_{\Lambda},H]=U^h[q_{\Lambda},P](U^h)^{-1}={\cal O}(h^\infty)$$ 
so that (i) holds for $Q_{\Lambda}$ (for simplicity, we still denote by $(\varphi,\iota)$ the set $(\varphi',\iota')$ 
of action-angle variables given in Theorem 1.1).

We can take $b\in\widetilde S^0(\Sigma_4^{\delta})$ of the 
form $\beta({\iota_1\over h^{\delta}})$, with $\beta\in C^\infty({\bf R})$
supported in $\eta_1\leq-2$, as in Proposition A.3. Then $b$ and $Q_\Lambda$ have disjoint supports, 
and $BQ_\Lambda$ is negligible in the sense of Definition A.1. The same holds for $A(\Id-Q_\Lambda)$ and (ii) follows. 

(iii) follows from the functional calculus recalled in App.A.b, and  
the last statement from the fact that $Q_\Lambda^*$ has the same properties as $Q_\Lambda$.  $\clubsuit$
\smallskip
In particular, if $A(\varphi,\iota)=0$ in a sufficiently large $h^{\delta}$-neighbd of $\Lambda$, 
then for all $N>0$ there is $C_N>0$ such that~:
$$\|{i\over h}[A,Q_\Lambda]\|\leq C_N h^N \leqno(1.19)$$
Let now $\Sigma$ be separated by the tori $\Lambda_1=\Lambda_{\mu_1}$ and $\Lambda_2=\Lambda_{\mu_2}$, into $\Sigma_3$ and 
$\Sigma_4$ as in (1.13). Assume for simplicity that $\Sigma$ has no boundary, so that $\Sigma_3$ and $\Sigma_4$ are connected.
\medskip
\noindent {\bf Proposition 1.4}: {\it Under the hypothesis above, 
there exists $Q_3(x,\xi,h)\in S^0_{\delta}(M)$ (quasi-projector on $\Sigma_3$) verifying~:

\noindent (i) For all $N>0$, there exists $C_N>0$ such that~:
$$\|{i\over h}[Q_3,H]\|\leq C_{N}h^N \leqno(1.22)$$
\noindent (ii) Let $a\in S^0(F_3^{\delta/2})$, $b\in S^0(F_4^{\delta/2})$ be supported outside a sufficiently large $h^{\delta/2}$-neighborhood 
of $\Lambda_1$ and $\Lambda_2$, and $A,B\in L^0_{\delta/2}(M)$ the corresponding operators.
Then for all $N>0$, there exists $C_N>0$ such that~:
$$\|A(\Id-Q_3)\|\leq C_{N}h^N, \quad  \|BQ_3\|\leq C_{N}h^N\leqno(1.23)$$
\noindent (iii) If $A\in L^0_0(M)$, we have 
$$\|{i\over h}[A,Q_\Lambda]\|\leq Ch^{-\delta}\leqno(1.24)$$
Properties (1.22-24) hold for $Q_3^*$ as well. }
\smallskip
\noindent {\it Proof}: 
Let $\widetilde\chi_j\in C^\infty_0(T^*M)$, $j=1,2$, equal to 1 near $\Lambda_j$, and for $\chi_j$ as in (1.18),
$\chi_j\equiv1$ on supp$\widetilde\chi_j$, and also
$\widetilde\chi_3\in C^\infty_0(T^*M)$ be equal to 1 on $F_3\setminus\neigh (\Lambda_1\cup\Lambda_2)$ in $T^*M$, 
such that 
$\widetilde\chi_1+\widetilde\chi_3+\widetilde\chi_2=1$ on $F_3$. 
We glue the $Q_{\Lambda_i}$'s (after undoing the (local) canonical transformations $\kappa_0$ that take local coordinates $(x,\xi)$ in $M$
to action-angle coordinates $(\varphi,\iota)$, and the corresponding FIO). Consider the $h$-PDO
$$Q_3=Q_{\Lambda_1}\widetilde\chi_1^w(x,hD_x)+\widetilde\chi_3^w(x,hD_x)+Q_{\Lambda_2}\widetilde\chi_2^w(x,hD_x)$$ 
with symbol in $S^0_{\delta}(M)$, it also satisfies conclusions (i)-(ii) of Lemma 1.3. Namely
$$[H,Q_3]= Q_{\Lambda_1}[H,\widetilde\chi_1^w]+Q_{\Lambda_2}[H,\widetilde\chi_2^w]+
[H,\widetilde\chi_3^w]+[H,Q_{\Lambda_1}]\widetilde\chi_1^w+[H,Q_{\Lambda_2}]\widetilde\chi_2^w$$
The last 2 terms on the RHS are ${\cal O}(h^\infty)$ by (i), while for $j=1,2$,
$Q_{\Lambda_j}[H,\widetilde\chi_j^w]=[H,\widetilde\chi_j^w]$ on $\Sigma_3$
mod ${\cal O}(h^\infty)$, so the  
the sum of the 3 first terms vanishes mod ${\cal O}(h^\infty)$ because of $\widetilde\chi_1+\widetilde\chi_3+\widetilde\chi_2=1$. 
Properties (1.23) and (1.24) are derived similarly. $\clubsuit$
\bigskip
\noindent {\bf 2. Proof of Theorem 0.1}.
\smallskip
Proof of Thm 0.1 is very close to [Sh], but makes use of convergence in the mean of Wigner measures.
We recall first some well-known facts
about microlocal semi-classical spectral asymptotics, and refer to [Iv] for details.
We conclude the proof as in [Sh] by a dichotomy argument using the existence of symmetric, disjoint invariant tori.
\smallskip
\noindent{\it a) Semi-classical trace formulas and Wigner measures.}
\smallskip
Assume $\Sigma=\{H_0(x,\xi)=E\}$ is non critical for $H_0$, and $\Sigma_E$ is compact. 
Let $dL_E(x,\xi)$ be the (normalized) Liouville measure on $\Sigma_E$, i.e. 
$$dL_E(x,\xi)=\bigl(\int_{H_0(x,\xi)=E}{d\sigma_E\over|\nabla H_0|}\bigr)^{-1}{d\sigma_E(x,\xi)\over|\nabla H_0(x,\xi)|}$$
where $d\sigma_E(x,\xi)$ is the surface measure on $\Sigma_E$.

Assume $H^w(x,hD_x;h)$ has only discrete spectrum near $E$, and
let $(\lambda_j(h))_{j\geq0}$ be the sequence of its eigenvalues, counted with multiplicity, in the energy window
$I(h)=[E-h^\delta,E+h^\delta]$ for some $0<\delta<1$,
$(u_j(h))_{j\geq0}$ the corresponding sequence of normalized eigenfunctions,
and $J(h)=\{j\in{\bf N}: \lambda_j(h)\in I(h)\}$, so that $|J(h)|={\cal O}(h^{\delta-2})$. 
The next result follows easily from Weyl asymptotics: see
[PeR,Remark 5.2] and [HeMaR], based on earlier ideas of Ivrii, for details and more advanced results. 
\medskip
\noindent {\bf Proposition 2.1}: {\it Let $A=a^w(x,hD_x)$, with $a\in S^0(M)$. Under assumptions above, we have}:
$$\lim_{h\to0}{1\over |J(h)|}\Sum_{j\in J(h)}\bigl(a^w(x,hD_x)u_j(h)|u_j(h)\bigr)=\int_{\Sigma_E}a(x,\xi)dL_E(x,\xi)\leqno(2.1)$$
\smallskip
We change Weyl quantization $a^w(x,hD_x)$ to anti-Wick quantization $a^{\overline w}(x,hD_x)$, in order 
to preserve positivity of observables $a(x,\xi)$. 
This change of quantization only modifies $a^w(x,hD_x)$ by a compact operator, of norm ${\cal O}(h)$. Recall (see e.g. [HeMaRo])
$a^{\overline w}(x,hD_x)=b^w(x,hD_x,h)$ with 
$$b(x,\xi,h)=(\pi h)^{-d}\int\int e^{-\bigl((x-y)^2+(\xi-\eta)^2\bigr)/h}a(y,\eta) \ dyd\eta$$
With this choice, $a\mapsto\bigl(a^{\overline w}(x,hD_x)u_j(h)|u_j(h)\bigr)$ defines a positive linear form on $C_0^\infty(T^*M)$,
and there is a probability measure $d\mu_j^h$ on $T^*M$ such that 
$$\bigl(a^{\overline w}(x,hD_x)u_j(h)|u_j(h)\bigr)=\int a(x,\xi)d\mu_j^h(x,\xi)\leqno(2.2)$$
It follows then from Proposition 2.1 that
$$\lim_{h\to0}{1\over |J(h)|}\Sum_{j\in J(h)}d\mu_j^h(x,\xi)=dL_E(x,\xi)\leqno(2.3)$$
in the sens of vague convergence of Radon measures on $T^*M$. We can also write $d\mu_j^h=W^hu_j(h)\,dx\,d\xi$, 
where $W^h$ is the Wigner transformation
$$W^hu_j(x,\xi;h)={1\over(2\pi)^d}\int e^{iy\xi}u_j(x-h{y\over2};h)\overline {u_j}(x+h{y\over2};h)\, dy$$
(in local cordinates) and call $d\mu_j^h$ a Wigner measure~; 
its limit points as $h\to0$ are the {\it semi-classical measures} of $u_j(h)$~; see [HeMaRo], [GeLe]
and references therein. We are interested in their mean (2.3). 
\medskip
\noindent {\bf Proposition 2.2}: {\it For any $a\in S^0(F_3^{\delta/2})$, there is $a_0(x,\xi)\in L^\infty(M)$, supported in $F_3$ such that:}
$$\lim_{h\to0}{1\over |J(h)|}\Sum_{j\in J(h)}W^hu_j(x,\xi;h)a(x,\xi;h)\,dx\,d\xi=
\int_{\Sigma_E}a_0(x,\xi)dL_E(x,\xi)\leqno(2.4)$$
\smallskip
\noindent{\it Sketch of the proof}:
One has to complete the argument leading to Proposition 2.1 in the spirit of [FeGe]. Let ${\cal I}={\cal I}_2\cup{\cal I}_2$.
Consider the bundle $N({\cal I})$,
normal to ${\cal I}$ with fibers $N_\rho({\cal I})=T_\rho(T^*M)/T_\rho({\cal I})$, and its compactification $\overline N({\cal I})$
obtained by adding a sphere at infinity. 
Adapting the proof of [FeGe] in the context of Proposition 1.2, we see that 
concentration of ${1\over |J(h)|}\Sum_{j\in J(h)}d\mu_j^h(x,\xi)$, at scales 1 and $h^{\delta/2}$ along ${\cal I}$,
is described by the sum of
$dL_E(x,\xi)$, and a ``2-scaled'' measure $d\nu_{\cal I}(x,\xi)$ supported on $\overline N({\cal I})$. 
Since we restrict to observables supported in $F_3$, $\nu_{\cal I}$ doesn't contribute, and we are left with
$dL_E(x,\xi)$. Following [FeGe,Sect.2] we find that, for such an $a$,  the LHS of (2.4) converges, by dominated convergence theorem, to 
$\int\lim_{h\to0}\alpha({\iota_1\over h^{\delta}})\chi({\iota\over h^{\delta}})dL_E(x,\xi)=\int a_0(x,\xi)dL_E(x,\xi)$, where $a_0\in L^\infty(M)$ 
is supported on $F_3$. This yields easily (2.4). $\clubsuit$
\medskip
Let $Q=Q_3$ be the quasi-projector constructed in Proposition 1.4, and
$a\in S^0(F_3^{\delta/2})$, $a\geq0$ (near $\Lambda_1$, we take as before $a$
of the form $\alpha({\iota_1\over h^{\delta}})b(x,\xi;h)$.~) By functional calculus reviewed in App.A.b, we can define $Q^*a^wQ$
as an operator in $L^0_{\delta/2}(M)$, whose symbol is again in $S^0(F_3^{\delta/2})$. Following
the proof of Proposition 2.1, we have
$$\lim_{h\to0}{1\over |J(h)|}\Sum_{j\in J(h)}\bigl(Q^*a^w(x,hD_x)Q u_j(h)|u_j(h)\bigr)=
\int_{\Sigma_E}(Q^*aQ)_0(x,\xi)dL_E(x,\xi)\leqno(2.6)$$
where $(Q^*aQ)_0\in L^\infty(M)$ has support in $F_3$. 
Passing again to anti-Wick quantization gives
$$\lim_{h\to0}{1\over |J(h)|}\Sum_{j\in J(h)}\int(Q^*aQ)(x,\xi;h)d\mu_j^h(x,\xi)=\int_{\Sigma_E}(Q^*aQ)_0(x,\xi)dL_E(x,\xi)\leqno(2.7)$$
in the sens of vague convergence of Radon measures on $T^*M$. 
\medskip
\noindent{\it b) A pair of orthogonal quasi-modes; end of the proof of Thm 0.1 using symmetry}.
\smallskip
The space $\Ran Q_3\subset L^2(M)$ is a quasi-mode supported microlocally in the component $\Sigma_3$ of the energy surface between $\Lambda_1$
and $\Lambda_2$. Similarly, $\Ran (\id -Q_3)\subset L^2(M)$ defines another quasi-mode supported on its complement $\Sigma_4$.
As was already observed in [CdV2] in the case of a quasi-integrable Hamiltonian ${\cal H}_\lambda$,
quasi-modes supported on different connected components of $\Sigma$ between KAM tori are
orthogonal. Here we show that such a conclusion holds in case of Maupertuis-Jacobi correspondence, up to extraction of a subsequence. 

Let $v_j(h)=Qu_j(h)$,  $w_j(h)=(\id -Q)u_j(h)$, where $Q=Q_3$. 
By Proposition 1.4 we have
$$(H-\lambda_j)v_j(h)=[H,Q]u_j(h)+Q(H-\lambda_j)u_j(h)={\cal O}(h^\infty)$$ 
in $L^2(M)$, and similarly for $w_j(h)$. As in [Sh], consider the Gram-Schmidt orthonormalization of $v_j(h)$ and $w_j(h)$~:
$$v'_j(h)={v_j(h)\over\|v_j(h)\|}, \quad w'_j(h)={w_j(h)-(v'_j(h)|w_j(h))v'_j(h)\over\|w_j(h)-(v'_j(h)|w_j(h))v'_j(h)\|}\leqno(2.8)$$
They provide a quasi-mode for $H$, when $j$ runs over some subset of $J(h)$, if the denominators are bounded from below. 
These quasi-modes may be associated with other invariant tori situated between the $\Lambda_j$'s, or tori of lower dimension, or other invariant 
subsets in the energy surface $\Sigma_E$. As in [Sh] we have~:
\medskip
\noindent{\bf Lemma 2.2}: {\it Let $a\in S^0(F_3^{\delta/2})$, $b\in S^0(F_4^{\delta/2})$. Let $u_j(h)$ be a normalized 
eigenfunction of $H$ with eigenvalue $\lambda_j(h)$, and assume 
$$|\int_{F_3^{\delta/2}} a(x,\xi)d\mu_j^h(x,\xi)|\geq c>0, \quad |\int_{F_4^{\delta/2}} b(x,\xi)d\mu_j^h(x,\xi)|\geq c>0\leqno(2.9)$$
uniformly for $h>0$ small enough. Then~:}
$$\leqalignno{
&c_1\leq \|v_j(h)\|\leq c_1^{-1}, \quad c_2\leq \|w_j(h)\|\leq c_2^{-1} \ \hbox{for some} \ c_1,c_2>0&(2.10)\cr
&|(v_j(h)|w_j(h))|\leq c_3 \|v_j(h)\|\|w_j(h)\|, \ 0<c_3<1&(2.11)\cr
}$$
And there follows the~:
\smallskip
\noindent {\bf Corollary 2.3}: {\it Assume that the conditions of Lemma 2.2 are fulfilled for some subsequence $j_k$, then we may find a 
sequence of pairs of
orthonormal functions $(v'_{j_k},w'_{j_k})$ such that}
$$\|(H^w-\lambda_{j_k})v'_{j_k}\|={\cal O}(h^\infty), \quad \|(H^w-\lambda_{j_k})w'_{j_k}\|={\cal O}(h^\infty)$$
\smallskip
This means that the sequence of $\lambda_{j_k}(h)$ is asymptotically degenerated. 
We are left to show that the presence of a 
symmetry for ${\cal H}={\cal H}_\lambda$ guarantees the conditions of Lemma 2.2. Since ${\cal H}(x,\xi)$ is invariant under 
$\Gamma:(x,\xi)\mapsto(x,-\xi)$, and
because of MJC, the energy surface $\Sigma=\Sigma_E$ of $H$ is also invariant for $\Gamma$. 
As in [Sh], we start with~:
\smallskip
\noindent {\bf Lemma 2.4}: {\it Let $a\in C^\infty({\cal V})$ odd under $\Gamma$, i.e. $a(x,\xi)=-a(x,-\xi)$. 
Then the semi-classical measures 
$\mu_j^h(x,\xi)$ are asymptotically invariant with respect to $\Gamma$ in the mean, i.e.} 
$$\lim_{h\to0}{1\over |J(h)|}\Sum_{j\in J(h)}\bigl(a^{\overline w}(x,hD_x;h)u_j(h)|u_j(h)\bigr)=0\leqno(2.12)$$
\smallskip
Assume now we are given 4 invariant tori (separated in phase-space) 
with Diophantine frequency vectors, $\Lambda_1$, $\Lambda_2=\Gamma\Lambda_1$, $\Lambda_3$, and $\Lambda_4=\Gamma\Lambda_3$
in $\Sigma$. The tori $\Lambda_1$ and $\Lambda_2$ divide $\Sigma$ into 2 domains, denote them by $\Sigma_3$ and $\Sigma_4$~; 
in the same way, the tori $\Lambda_3$ and $\Lambda_4$ divide $\Sigma$ into 2 domains, and we
denote them by $\Sigma_1$ and $\Sigma_2$. The numerotation is chosen in such a 
way that  $\Lambda_i\subset\Sigma_i$, $i=1,\cdots,4$. We define $F_1$ and $F_2$ accordingly.
Let also $\Lambda_i^{\delta }$ be a $h^{\delta/2}$-nghbd of $\Lambda_i$.

We proceed by dichotomy, dividing the sequence $u_j(h)$ of eigenfunctions of $H$ in $I(h)$ into 2 sets $S_1(h)$ and 
$S_2(h)$. Let $\e _0>0$ to be chosen small enough.
We say that eigenfunction $u_j(h)$ belongs to $S_1(h)$ iff $\mu_j^h(\Lambda_1^{\delta/2})>\e _0$ and $\mu_j^h(\Lambda_2^{\delta/2})>\e _0$.
Otherwise it belongs to $S_2(h)$. Now we choose $0\leq a_i\in C^\infty_0(F_i^{\delta/2}))$, $i=1,\cdots,4$, such that $a_1\equiv1$ on 
$\Lambda_1^{\delta/2}$, $a_2\equiv1$ on $\Lambda_2^{\delta/2}$, $a_3\equiv1$ on $F_3^{\delta/2}\setminus(\Lambda_1^{\delta/2}\cup\Lambda_2^{\delta/2})$, 
and $a_4\equiv1$ on $F_4^{\delta/2}\setminus(\Lambda_1^{\delta/2}\cup\Lambda_2^{\delta/2})$. Let also $a_i$ be chosen in such a way that the symmetries 
$a_1=a_2\circ \Gamma$, $a_3=a_4\circ \Gamma$ hold. We consider also the anti-Wick quantization $a_i^{\overline w}(x,hD_x;h)$ of $a_i$ as before.

By Lemma 2.5 and $a_1=a_2\circ \Gamma$ we have: 
$\lim_{h\to0}{1\over |J(h)|}\Sum_{j\in J(h)}\int_{\cal V} (a_1-a_2)d\mu_j^h=0$, while 
$$\lim_{h\to0}{1\over |J(h)|}\Sum_{j\in J(h)}\int_{\cal V} (a_1+a_2)d\mu_j^h=\int_{\Sigma}(a_1+a_2)dL_E=2C_0>0$$ 
so
$$\lim_{h\to0}{1\over |J(h)|}\Sum_{j\in J(h)}\int_{F_1} a_1 d\mu_j^h=
\lim_{h\to0}{1\over |J(h)|}\Sum_{j\in J(h)}\int_{F_2} a_2 d\mu_j^h=C_0>0$$ 
Since $a_i\geq0$, $i=1,2$, there is a subsequence $J_i(h)\subset J(h)$ of relative density 1, and $c>0$ such that
for all $j\in J_i(h)$, $\int_{F_i} a_i(x,\xi)d\mu_j^h(x,\xi)\geq c$, uniformly as $h>0$ small enough. 
This holds for all 
$j\in J_{12}(h)=J_1(h)\cap J_2(h)$ which is again of relative density 1 [GeLe,Lemma 5.1, and its proof].

The same argument yields 
$$|\int_{F_3} a_3(x,\xi)d\mu_j^h(x,\xi)|\geq c, \quad 
|\int_{F_4} a_4(x,\xi)d\mu_j^h(x,\xi)|\geq c$$
when
$j\in J_{34}(h)$, a sequence of relative density 1. So taking $J'(h)=J_{12}(h)\cap J_{34}(h)$,
we have $\int_{F_i} a_i(x,\xi)d\mu_j^h(x,\xi)\geq c$ for $i=1,\cdots,4$, uniformly for $j\in J'(h)$ and $h>0$ small enough.

In the definition of $S_1(h),S_2(h)$, take $\e _0=c$. 
Assume first that $S_1(h)\cap J'(h)$ is of relative density 1, then the conditions of Lemma 2.2 are satisfied for the domains
$\Sigma_1$, $\Sigma_2$, and observables $a_1$, $a_2$, because $\int_{F_i} a_i(x,\xi)d\mu_j^h(x,\xi)\geq \e _0=c$, for $i=1,2$.

Otherwise, $S_2(h)\cap J'(h))$ will be of relative density 1, and the conditions of Lemma 2.2 are satisfied for the domains
$\Sigma_3$, $\Sigma_4$, and observables $a_3$, $a_4$, because $\mu_j^h(\Sigma^\delta)=1$ and
$\int_{F_i} a_i(x,\xi)d\mu_j^h(x,\xi)\geq c'$, for $i=3,4$.

Applying Corollary 2.4, we have found in both cases a subsequence of relative density 1 of splitted, (or asymptotically degenerated) eigenvalues.
This completes the proof of Theorem 0.1. $\clubsuit$.
\bigskip
\noindent{\bf 3) Quasi-modes for rational Lagrangian tori and Larmor precession}.
\medskip
Let ${\cal H}$ be completely integrable. 
We assume here that $\Lambda$ is a Lagrangian torus in the energy shell $\{{\cal H}={\cal E}\}=\{H=E\}$ 
invariant under the flow of $X_{\cal H}$, which is conjugated with a linear with rational frequency
vector $\widetilde\omega$ on the torus. Making a linear transformation $T\in{\SL}_2({\bf Z})$ on the angles, we can assume that
$\widetilde\omega=(\widetilde\omega_1,0)$. MJC induces a reparametrization of time, of the 
form $dt={\cal G}(\varphi_1+\widetilde\omega_1\tau,\varphi_2)d\tau$. Here we have set $\varphi=(\varphi_1,\varphi_2)\in{\bf T}^2$,
but the constructions below carry to the case where $\varphi\in{\bf T}^d$.

Following the proof of [DoRo,Theorem 0.4], but using only Fourier expansion in the $\varphi_1$ variable, 
we can show that there is a reparametrization of ${\bf T}^2$, of the form 
$$\Phi : \varphi\mapsto\psi=\Phi(\varphi)=\varphi+\widetilde\omega g(\varphi)$$ 
where $g$ is a smooth, scalar periodic function (in particular, $\psi_2=\varphi_2$.~)
The motion on $\Lambda$ induced by this reparametrization is periodic with frequency vector $\omega=(\omega_1,\omega_2)=(\omega_1,0)$, 
$$\omega_1=\omega_1(\psi_2)={\widetilde\omega_1\over\langle {\cal G}\rangle_{\psi_2}}$$ 
where $\langle {\cal G}\rangle_{\psi_2}$ denotes the average with respect to $\psi_1$. It is a smooth periodic
function of $\psi_2=\varphi_2\in{\bf T}$. 
Near $\Lambda$ we apply Darboux-Weinstein theorem in the special form given in [BeDoMa,Theorem 2], [DoRo,Theorem 1.4], thus we can write 
Hamiltonian $H$ in some symplectic action-angle coordinates $(x,\xi)$ where $x=(x_1,x_2)\in{\bf T}^2$ 
stands for $\psi$ above, $\xi=(\xi_1,\xi_2)\in{\bf R}^2$ the dual coordinate and $\Lambda$ is given 
by $\xi=0$. This gives, with $H_0=H|_\Lambda$~:
$$H=H_0+\omega_1(x_2)\xi_1+a(x,\xi)\leqno(3.1)$$
where $a(x,\xi)={\cal O}(|\xi|^2)$ is a smooth periodic function on $x\in{\bf T}^2$.
This reminds of Larmor precession in a magnetic field for the motion of a particle taking place on a torus, see [ArKoNe,Sect.6.4]. 
In this model the  ``fast variable'' $x_1$ stands for the direction of the ``unperturbed
orbits'' (the small circles), and the ``slow variable'' $x_2$ for the direction of the ``drift''.

We use the method of averaging in a single variable. Consider a generating function of the form
$$\widetilde S(x,\eta)=\langle x,\eta\rangle+S(x,\eta), \quad S(x,\eta)={\cal O}(\eta^2)\leqno(3.2)$$
Writing $S(x,\eta)=\eta_1^2S_{11}(x)+2\eta_1\eta_2S_{12}(x)+\eta_2^2S_{22}(x)+\cdots$, we substitute
$\xi=\eta+{\partial S\over\partial x}(x,\eta)$, $y=x+{\partial S\over\partial \eta}(x,\eta)$
in (3.1) so that, with $a(x,\xi)=a_{11}(x)\xi_1^2+2a_{12}(x)\xi_1\xi_2+a_{22}(x)\xi_2^2+\cdots$
$$\eqalign{
H=H_0&+\omega_1(x_2)\eta_1+\eta_1^2\bigl(\omega_1(x_2){\partial S_{11}\over\partial x_1}(x)+a_{11}(x)\bigr)+
2\eta_1\eta_2\bigl(\omega_1(x_2){\partial S_{12}\over\partial x_1}(x)+a_{12}(x)\bigr)+\cr
&+\eta_2^2\bigl(\omega_1(x_2){\partial S_{22}\over\partial x_1}(x)+a_{22}(x)\bigr)+\cdots\cr
}$$
Let $b_{ij}(x_2)=\langle a_{ij}(x)\rangle_{x_2}$ be the average of $a_{ij}$ with respect to the ``fast variable'' $x_1\in{\bf T}$,
$x_2\in{\bf T}$ being held fixed.
Then we can solve the equations
$$\omega_1(x_2){\partial S_{ij}\over\partial x_1}(x)+a_{ij}(x)=b_{ij}(x_2)$$
leading to 
$$H(x_2,\xi)=H_0+\omega_1(x_2)\eta_1+\eta_1^2b_{11}(x_2)+2\eta_1\eta_2b_{12}(x_2)+\eta_2^2b_{22}(x_2)+{\cal O}(|\eta|^3)$$
This can be done to all orders in $\eta$. So for each $N$, we have the decomposition, writing again $\xi$ instead of $\eta$
$$H(x_2,\xi)=H_0+\omega_1(x_2)\xi_1+H_N(x_2,\xi)+{\cal O}(|\xi|^{N+1})\leqno(3.3)$$
where $H_N(x_2,\xi)$ is a polynomial in $\xi$ of degree $N$ and vanishing of order 2 at $\xi=0$. 

The point is that 
$H'_N(x_2,\xi)=\omega_1(x_2)\xi_1+H_N(x_2,\xi)$ is in involution with $\xi_1$, allowing for separation of variables microlocally near $\xi=0$.
We make the non degeneracy hypothesis
$${\partial^2 H'_N\over\partial \xi_2^2}(x_2,0,0)\neq0\leqno(3.5)$$ 
First we consider the set of points in the characteristic variety where
the momentum map is regular, i.e. when $dH'_N$ is not parallel to $d\xi_1$. 
The component of $dH'_N$ along $dx_2$ is $\omega'_1(x_2)\xi_1+{\partial H_N\over\partial x_2}(x_2,\xi)$, 
while those along $d\xi_1$ and $d\xi_2$ are respectively $\omega_1(x_2)+{\partial H_N\over\partial \xi_1}(x_2,\xi)$ and 
${\partial H_N\over\partial \xi_2}(x_2,\xi)$. 
When $\xi_1=0$, ${\partial H_N\over\partial x_2}(x_2,0,\xi_2)={\cal O}(\xi_2^2)$, while 
${1\over\xi_2}{\partial H_N\over\partial \xi_2}(x_2,0,\xi_2)\neq0$,
so $dH'_N$ can only be parallel to $d\xi_1$ on $\xi=0$. 

Then we look at the set of points where the momentum map is singular, i.e. at $\Sigma_1=\{dH'_N\parallel d\xi_1\}$. 
For such a point, we have 
$$\omega'_1(x_2)\xi_1+{\partial H_N\over\partial x_2}(x_2,\xi)=0, \quad {\partial H'_N\over\partial \xi_2}(x_2,\xi)=0$$
So if $\omega'_1(x_2)\neq0$, we need $\xi_1=0$,
and ${\partial H'_N\over\partial x_2}(x_2,0,\xi_2)={\partial H'_N\over\partial \xi_2}(x_2,0,\xi_2)=0$. When (3.5) holds, 
the only critical point near $\xi_2=0$ of $\xi_2\mapsto H'_N(x_2,0,\xi_2)$ 
is $\xi_2=0$, and $\Sigma_1=\{{\partial H'_N\over\partial x_2}(x_2,0)={\partial H'_N\over\partial \xi_2}(x_2,0)=0\}$. 
Under (3.5), it is easy to check that 
$d{\partial H'_N\over\partial x_2}(x_2,0,0)$ and $d{\partial H'_N\over\partial \xi_2}(x_2,0,0)$ are not parallel, so 
the critical set $\Sigma_1$ given by $\xi_1=\xi_2=0$ 
has codimension 2 in the parameter space $x_2,\xi_1,\xi_2$ (the variable $x_1$ is cyclic). 

At last, if $\omega'_1(x_2)=0$ at some point $x_2=x_2^0$, the critical set $\Sigma_1$ is given (locally) by 
$\omega'_1(x_2)\xi_1+{\partial H_N\over\partial x_2}(x_2,\xi)=0$, ${\partial H'_N\over\partial \xi_2}(x_2,\xi)=0$ 
Under (3.5), the second equation gives $\xi_2=\Xi_2(\xi_1,x_2)={\cal O}(\xi_1)$, and substituting into the first one we get
$\omega'_1(x_2)+\Xi_1(\xi_1,x_2)=0$, with $\Xi_1(\xi_1,x_2)={\cal O}(\xi_1)$. Assuming further that $\omega''_1(x_2^0)\neq0$, 
we find under (3.5) that 
$d{\partial H'_N\over\partial x_2}(x_2,0,0)$ and $d{\partial H'_N\over\partial \xi_2}(x_2,0,0)$ are not parallel, so again
the critical set $\Sigma_1$ given by $\xi_1=\xi_2=0$ 
has codimension 2 in the parameter space $x_2,\xi_1,\xi_2$ near $x_2^0$.

So in any case $\Sigma_1$ is a smooth submanifold of 
codimension 2 under the non degeneracy hypothesis (3.5) and that $\omega''(x_2)\neq0$ on $\omega'_1(x_2)=0$,
i.e. when $\omega_2$ is a Morse function.

To the locus $\omega'_1(x_2)=0$
correspond periodic trajectories for $H'_N$ (Larmor circles which are not drifting), which are separatrices between 
domains of the energy surface $E=0$. For a global energy picture of Hamiltonian systems in involution and their Reeb graphs, see e.g. [BrDoNe].
 
Let us find approximate eigenfunctions for $H'_N(x_2,hD_x)$ under (3.5) and the hypothesis that $\omega_2(x_2)$ is a Morse function.
Because $[hD_{x_1},H'_N(x_2,hD_x)]=0$, we consider the joint spectrum, and 
eigenfunctions of the form $u_1(x_1,h)\otimes u_2(x_2,h)$. 
We may consider general Floquet periodic eigenfuntions $u_1(x_1,h)$ on ${\bf T}$, and adapt the microlocal Floquet-Bloch theory
of Appendix A to this situation, but for the sake of simplicity, we restrict here 
to eigenfunction for $hD_{x_1}$ with periodic boundary condition 
of the form $e^{ik_1x_1/h}$ with $k_1\in2\pi h{\bf Z}$. 

Substituting into $H'_N(x_2,hD_x)$ we find 
$$H'_Nu(x,h)=e^{ik_1x_1/h}\bigl(\omega_1(x_2)k_1+H_N(x_2,k_1,hD_{x_2})\bigr)u_2(x_2,h)$$
To fix the ideas we consider the particular case where $H_N(x_2,\xi)={1\over2}\xi_2^2$, so we get the Schr\"odinger operator
$$P(x_2,hD_{x_2})={1\over2}(hD_{x_2})^2+\omega_1(x_2)k_1\leqno(3.6)$$ 
on $L^2({\bf T})$ with potential $k_1\omega_1(x_2)$. 
Because of harmonic approximation of the Hamiltonian near a non degenerate critical point [HeSj2], this does not restrict the generality. 
Since $\omega_1>0$, the situation differs according to the sign of $k_1$. 

If $0<k_1={\cal O}(h^\delta)$, $P(x_2,hD_{x_2})$
will have bound states of energy $E$ when $E\geq \Const k_1$, corresponding to a classical motion in potential wells close to the 
minima of $\omega_1$, with momentum $\xi_2={\cal O}(\sqrt{k_1})$. 
The spectrum on $P$ near $E$ will be the union of spectra of localized operators 
in each well. This means that the particle on the torus chooses regions where $\omega_1$ is small.

On the other hand, when $0<-k_1={\cal O}(h^\delta)$, $P(x_2,hD_{x_2})$
will have bound states of energy $E$ when $E\leq -\Const k_1$,
corresponding to a classical motion bouncing between the 
maxima of $\omega_1$. In this case, the particle with negative momentum $k_1$, will have to ``move up the stream'',
in a direction opposite to this of the Hamilton flow. This dynamics is unstable, and includes tunneling between successive maxima 
of $\omega_1$ on ${\bf T}$. 

Thus for simplicity we focus on $k_1>0$, and 
assume that $\omega_1(x_2)$ is a regular Morse function, taking its minimum value $\omega_0>0$ on a discrete set of nondegenerate
critical points on ${\bf T}$. So we look at the spectrum of (3.6) near a local non degenerate minimum $x_2^0$ of $\omega_1$ and for notational 
simplicity, we assume $x_2^0=0$. This amounts to consider quasi-modes supported near a Larmor circle which is not drifting.

Recall from (A.5) the class of symbols $\widetilde S^m_{\delta}({\bf T}^2)$. Because we make the harmonic approximation of $P$ near a minimum of 
$\omega_1$, we need also to scale $x_2$ by an factor $h^{1/2}k_1^{-1/4}$, and so introduce an inhomogeneity between $\xi_1$ and $\xi_2$,
see Appendix A. 
Details of the contruction of quasi-modes for $P$ are given in [HeSj,Thm 3.7].  

Of course, the general situation when (3.5) doesn't necessary hold is much more complicated, and the dynamics induced by MJC near
a rational torus on $T^*{\bf T}^2$ is far from being integrable. In this chaotic landscape, Larmor circles which are not drifting 
and we have just described, appear as 
islands of stability. See e.g. [O-de-Al] for a discussion of semi-classical chaotic systems. 
\bigskip
\noindent{\bf 4) Aharonov-Bohm effect on the sphere and projectively equivalent Finsler structures}.
\medskip
Consider the Lagrangian on the unit sphere ${\bf S}^2$
$$L(x,\vec v)=\sqrt{{1\over2}\vec v^2}+\langle\vec A(x),\vec v\rangle\leqno(4.1)$$
where $\vec A(x)$ is the radially symmetric potential vector
$\vec A(x)=\alpha(-{\sin q_1\over\cos q_2}, {\cos q_1\over\cos q_2}, 0)$
in equatorial coordinates $(q_1,q_2)$ where
$x_1=\cos q_1\cos q_2, x_2=\sin q_1\cos q_2, x_3=\sin q_2$.
This is a smooth vector field outside the poles $q_2=\pm{\pi\over2}$, verifying $d\vec A(x)=0$ 
(the covariant derivative being taken with respect to the standard metric on $M$, induced by the Euclidean metric on ${\bf R}^3$).
In these coordinates, 
$$L(x,v)=\widetilde L(q,\dot q)=\sqrt{{1\over 2}((\cos q_2)^2\dot q_1^2+\dot q_2^2)}+\alpha \dot q_1\leqno(4.2)$$
Physically, Lagrangian $L$ involves a thread of magnetic flux through the poles, with strength
$\alpha=\oint \langle\vec A,\,d\vec \ell\rangle$  (the circulation of $\vec A$ on a loop encircling the poles).
We call (4.1) ``Aharonov-Bohm Lagrangian'' on ${\bf S}^2$, where we have replaced the kinetic energy by the length functional, see [Rui].

When $|\alpha|<1$, (4.1) or (4.2) also define a Finsler metric on $M={\bf S}^2$. 
A Finsler metric on the manifold $M$ is a smooth positive function on $TM\setminus0$ enjoying the properties of {\it homogeneity}~:
$F(x,\lambda v)=\lambda F(x,v), \forall v\in T_xM\setminus0, \forall \lambda\in]0,\infty[$, 
and {\it strong convexity}, i.e. if we set $f(x,v)={1\over2}F(x,v)^2$, then $D^2_vf(x,v)$ is positive definite. 
These metrics are not {\it reversible}, as soon as they contain a linear term.
Finsler metrics we consider here are a special class known as ``Randers metrics'', i.e. metrics of the form
$$F(x,v)=\sqrt{g_x(v,v)}+g_x(v,X)\leqno(4.3)$$ 
where $g_x$ is a Riemannian metric tensor on $M$ and $X$ a real vector field satisfying $g_x(X,X)<1$. 
With a Finsler metric, we associate a Hamiltonian by the usual prescription. In the case of 
a Randers metric, this yields a ``Randers symbol'' on $T^*M$, having the form~:
$$H(x,\xi)=\sqrt{\widetilde h_x(\xi,\xi)}+\widetilde h_x(Y,\xi)=\lambda(x,\xi)+\eta(x,\xi)\leqno(4.4)$$
where $\widetilde h_x$ is a positive definite quadratic form on $T_x^*M$ and $Y$ a real vector field on $M$, satisfying 
$\widetilde h_x(Y,Y)<1$. If $X\neq0$, we emphasize that $\widetilde h_x$ is not the form on $T^*M$ dual to $g_x$. 
Finsler metrics play an important r\^ole in Geometrical Optics for inhomogeneous media (see [Du]).

The famous Katok example on the sphere ${\bf S}^2$ is constructed as follows. Let 
$g$ be the standard metric tensor 
on ${\bf S}^2$, and $Y_0\in T{\bf S}^2\setminus0$ the generator of a group of rotations
$R_0(t)$ of period $2\pi$.
We take $Y=\alpha Y_0$, $\alpha\in]-1,1[$, so that $g_x(Y,Y)<1$.
From the discussion above, it is clear that (4.1)-(4.2) and (4.3) define the same Finsler metric on ${\bf S}^2$, with $Y=\vec A$.
The geometry of Katok sphere is well-understood, see [Tay] and [Zi]. The Katok flow is integrable on ${\bf S}^2$, which follows from the 
fact that $\lambda(x,\xi)$ and $\eta(x,\xi)$ Poisson commute. Integrability holds in any dimension by a slightly more sophisticated argument.
In particular, when $\alpha$ is rational, the flow $\exp tH_\eta(x,\xi)$ is completely periodic on ${\bf S}^2$,
as is the geodesic flow on the standard sphere ${\bf S}^2$.
For irrational $\alpha$ instead, there are only 2 closed geodesics $\gamma_\pm$, both supported on the equator, but swept with different speeds
$1\pm\alpha$, due to the fact that the metric is not reversible. These orbits are disjoint in $T^*{\bf S}^2$.

The following property of {\it projectively equivalent} Finsler metrics, i.e. having same geodesics, is due to M.Hashigushi and 
Y.Ichijyo, see Example 3.3.2 in [ChSh]. Let $M$ be a manifold, $F(x,v)$ a Finsler metric, and $\beta$ a smooth 1-form on $M$, then 
$F+\beta$ is projectively equivalent to $F$ iff $\beta$ is closed. 
The 2 metrics on ${\bf S}^2$ (standard and Katok) verify these conditions, except for the fact that $\vec A$ is not smooth
at the poles~; the equator is the only geodesic they have in common, so they are not projectively equivalent. 

Projective equivalence extends Maupertuis-Jacobi correspondence in case of Finsler metrics. 
For irrational $\alpha$, standard and Katok metrics on ${\bf R}^2$ satisfy only ``partial'' MJC, in the sense that their 
only common Hamiltonian orbit is the (lift of) equator $\gamma$. We say that the singularity of $\vec A$ at the poles {\it breaks} MJC.
Recall from [Zi,p.145] the following fact: if  $\alpha$ is irrational, the periodic orbits $\gamma_\pm$ are critical points for 
the Lagrangian action $\int_I\widetilde L(q,\dot q)\, dt$ defined over all absolutely continuous loops $\gamma: I\to M$.
Moreover, Poincar\'e map ${\cal P}$ for $\gamma_\pm$ is tangent to rotations with angle 
$\beta_\pm={2\pi\over1\pm\alpha}$ respectively (with rescaled energy). In particular, $\gamma_\pm$ are
of elliptic type with irrational exponents, i.e. stable in the Hamiltonian sense.

Quantization of Finsler (Randers) symbols and wave kernels on Katok sphere are investigated in [Tay]. 
On the other hand, a general procedure for constructing a quasi-mode on a Riemannian manifold,
microlocalized near a Lagrangian manifold $\Lambda$ for the geodesic flow at some given energy $E$
has been devised in [We]; actually this quasi-mode is
associated with a fundamental cycle on $\Lambda$. On the standard sphere however, although the geodesic flow is integrable, 
the momentum map is singular on every closed geodesic: the sphere cotangent bundle $S^*{\bf S}^2$ is foliated by circles,
not by Lagrangian tori. 
Here we address the problem of constructing quasi-modes 
associated with either closed orbits $\gamma_\pm$ of Katok sphere, for the corresponding $h$-PDO (0.7) quantizing Hamiltonian (4.4),
$h$ being an extra parameter. By homogeneity, we can fix the energy level $E=1$.
This can be done since $\gamma_\pm$ are of elliptic type; in particular the argument set up in [Ral], using Gaussian beams,
can be easily extended to the case of a compact manifold, and we get the following:
\medskip
\noindent{\bf Theorem 4.1}: {\it Let $H(x,p)$ be Randers symbol associated with Bohm-Aharonov Lagrangian
$L(x,\vec v)=\sqrt{{1\over2}\vec v^2}+\langle\vec A(x),\vec v\rangle$ on $M={\bf S}^2$ as above, $A(x)=\alpha \dot q_1$
with irrational $|\alpha|<1$. Then for each of periodic orbits $\gamma=\gamma_\pm$, with rotation number $\beta=\beta_\pm$,
there is a quasi-mode of infinite order, i.e. 
a sequence of ``Planck constants'' $h_m>0$, indexed by $m=(m_1,m_2)\in{\bf Z}\times{\bf N}$, with $h_m\to0$ as $|m|\to \infty$, 
of normalized quasi-eigenfunctions $u_m=u_{h_m}\in L^2(M)$, and quasi-energies $E_m=E(h_m)\sim1+b_2h_m^2+\cdots$,
such that $(H(x,h_mD_x)-E_m)u_m={\cal O}(h_m^\infty)$. Moreover, the sequence $h_m$ is determined by the 
Bohr-Sommerfeld-Maslov quantization condition
$$C(h_m)=h_m(2\pi m_1+m_2\beta+{\beta\over2}+p\pi)$$
where $C(h_m)$ is the action integral $\int_\gamma p\, dx$, and $p\in{\bf Z}$ denotes Gelfand-Lidskii index for $\gamma$. }
\bigskip
\noindent {\bf Appendix A: Proof of Theorem 1.1}
\medskip
\noindent{\bf a) Wave-front sets and microlocalization}.
\smallskip
We recall here some facts from [Iv,Sect.1.3], and [Roy]. 
We consider families of objects ($L^2$ functions, bounded operators on $L^2$, or bounded sets in $T^*M$) depending 
on parameters, in particular on our ``Planck constant'' $0<h<h_0$. 
As {\it admissible} we consider {\it temperate functions}, i.e. functions $u=u_h\in h^{-m}L^2(M)$, or 
{\it temperate operators}, i.e. operators $A\in h^{-m}{\cal L}(L^2(M))$, for some $m>0$.
In particular, $h$-PDO's 
$$\widehat Au(x,h)=a^w(x,hD_x;h)u(x;h)=\int\int e^{i(x-y)\xi/h}a({x+y\over2},\xi;h)u(y)dyd\xi\leqno(A.1)$$ 
whose Weyl symbol $a(x,\xi;h)$ lies in ``H\"ormander class'' for some $0\leq\delta<1$
$$S^m_{\delta}(M)=\{a\in C^\infty(T^*M): |\partial_x^\alpha\partial_\xi^\beta a(x,\xi;h)|\leq C_{\alpha,\beta}h^{m-\delta(|\alpha|+|\beta|)/2}\}
\leqno(A.2)$$
(the condition $\delta\leq1$ being required to fulfill Heisenberg uncertainty principle) are {\it temperate operators}. 
In fact, $S^m_{\delta}(M)=h^mS^0_{\delta}(M)$.
We can as well use other standard quantizations of the symbol $a(x,\xi;h)$.

We can extend (A.2) by allowing anisotropies in $(x,\xi)$ variables, namely we introduce the class
$$S^m_{\delta,\nu,\gamma}(M)=\{a\in C^\infty(T^*M): |\partial_x^\alpha\partial_\xi^\beta a(x,\xi;h)|\leq C_{\alpha,\beta}h^m\nu^{-\beta}\gamma^{-\alpha}\}
\leqno(A.3)$$
for (variable) weights $\nu,\gamma\in{\bf R}_+$ such that $\inf _{1\leq j\leq d}(\nu_j\gamma_j)\geq h^\delta$. 

In the same way, $h$-FIO's associated with a non degenerate phase function and an amplitude in $S^m_{\delta,\nu,\gamma}(M)$ are admissible operators. 
Composition in the class of such admissible operators has natural properties, all stated in [Iv]. 

As {\it admissible} we consider {\it boxes} centered at some $\rho^0=(x^0,\xi^0)\in T^*M$, of the form
$$\Pi_{\rho^0}^{\delta,\nu,\gamma}=\{(x,\xi): |x_j-x_j^0|\leq\gamma_j, \ |\xi_j-\xi_j^0|\leq\nu_j\}$$
for weights $\nu,\gamma$ as above. When $\nu_j=\gamma_j=h^{\delta/2}$, we call $\Pi$ $\delta$-{\it isotropic}.
\medskip
\noindent{\bf Definition a.1}: {\it(i) Let $u$ be an admissible function, and $\Pi=\Pi_{\rho^0}^{\delta,\nu,\gamma}$ an admissible box. 
Then $u$ is {\it negligible} in 
$\Pi$ ($u\equiv0$ in $\Pi$) iff there exists an admissible observable ($h$-PDO) $a\in S^m_{\delta,\nu,\gamma}(M)$, such that $a=1$ in $\Pi$
and $Au(x,h)={\cal O}(h^\infty)$. We write $\rho^0\notin\WF u$, which defines a closed subset $\WF u\subset T^*M$ called the 
{\it wave-front set} or {\it oscillation front} in [Iv].

(ii) Similarly, let $A$ be an admissible operator, and $\Pi=\Pi'\times\Pi''$ an admissible box, centered in $({\rho'}^0,{\rho''}^0)$.
Then $A$ is negligible in $\Pi$ ($A\equiv0$ in $\Pi$) iff there exist admissible observable $a'\in S^m_{\delta',\nu',\gamma'}(M)$,
$a'=1$ in $\Pi'$, 
$a''\in S^m_{\delta'',\nu'',\gamma''}(M)$, $a''=1$ in $\Pi''$, and such that $\widehat A''\widehat A\widehat A'\equiv0$. 
We write $(\rho'^0,\rho''^0)\notin\WF A$, which defines a closed subset $\WF A\subset T^*M\times T^*M$.

(iii) When $\nu_j=\gamma_j=h^{\delta/2}$, we write simply $\WF ^\delta$ for $\WF$.}
\smallskip
Of course, we have the usual characterization of $\WF ^\delta$ using $h$-Fourier transforms, i.e. $\rho=(x_0,\xi_0)\notin \WF ^\delta u$
iff there is $\chi\in C^0_\infty(M)$ equal to 1 near $\rho$ such that 
${\cal F}_h(\chi\bigl({\cdot\over h^{\delta/2}})u\bigr)(\xi)={\cal O}(h^\infty)$ uniformly for $\xi$ in a $h^{\delta/2}$-neighbhd of $\xi_0$.

If $A$ is a $h$-PDO, $A$ is (pseudo) local, so we can choose $\Pi'=\Pi''$, and say $A$ is negligible in $\Pi'$.
while if $A$ is a $h$-FIO associated with the canonical
relation $\kappa$, then we can choose $\Pi'$ and $\Pi''$ so that their centers are related by $\rho''^0=\kappa(\rho'^0)$. 
Admissible (negligible) functions are Schwartz kernels of admissible (negligible) operators. 
\medskip
\noindent{\bf Definition a.2}: When $F$ is a (fixed) subset of $T^*M$ and
$0<\delta<1$, we denote by $F^\delta$ a $h^{\delta/2}$-nghbd of $F$. If $A$ is an admissible $h$-PDO, with symbol $a\in S^0_\delta(M)$
we say that its symbol belongs to
$S^0(F^{\delta})$, iff for all $\rho\notin F$, $\widehat A$ is negligible in the admissible box $\Pi'$ centered at $\rho$. 
\smallskip
\noindent{\it Example 1}: Let $\chi\in C^\infty_0(T^*M)$, $\chi(0,0)\neq0$~; then the symbol $\chi\bigl(h^{-\delta/2}(x-x^0,\xi-\xi^0)\bigr)$ 
defines an admissible $h$-PDO $\widehat A\in S^0(F^{\delta})$, with $F=\{\rho_0\}$.
\medskip
\noindent{\bf b) Pseudo-Differential calculus with periodic coefficients}.
\smallskip
When working in action-angle coordinates, locally $M={\bf T}^d$. 
The phase variable are $(x,\xi)=(\varphi,\iota)$, and it is convenient to take $\gamma_j=1$,
$\nu_j=h^\delta$ in (A.3). The corresponding class of symbols with periodic coefficients is given by: 
$$\widetilde S^m_{\delta}({\bf T}^d)=\{a\in C^\infty(T^*{\bf T}^d): |\partial_\varphi^\alpha\partial_\iota^\beta a(\varphi,\iota;h)|\leq 
C_{\alpha,\beta}h^{m-\delta|\beta|}\} \leqno(A.5)$$
In practice, $a(\varphi,\iota;h)$ is defined locally near $\iota=\iota_0$, which will be tacitely assumed in that definition.
One should keep in mind that when $\delta>0$,
the elements of $S^m_{\delta}({\bf T}^d)$ may not have well defined principal symbol, see e.g. [Roy].
Definition A.2 carries to $\widetilde S^0(F^{\delta})$ in the periodic case, but we keep in mind that if $F=F(\iota)\subset T^*M$, then
a $h^\delta$-neighbhd expressed in the $\iota$ variables alone stands for a $h^{\delta/2}$-neighbhd of $F$ in the $(x,\xi)$ variables.
We have the easy:
\medskip
\noindent{\bf Proposition b.1}: 
{Let $\chi\in C^\infty({\bf R}_+^d)$, vanishing in $\iota_1\leq\e _0$, $\e _0>0$~; then the symbol $\chi\bigl(h^{-\delta}\iota_1\bigr)$ 
lies in $\widetilde S^0(F^{\delta})$, with $F=\{\iota_1>0\}$.}
\smallskip
With a symbol $\widetilde S^m_{\delta}({\bf T}^d)$ we associate the operator $A:C^\infty({\bf T}^d)\to C^\infty({\bf T}^d)$ by the formula
$$Au(\varphi,\iota;h)=(2\pi)^{-d}\int_{{\bf T}^d}\, d\psi\Sum_{k\in{\bf Z}^d} e^{ik(\varphi-\psi)}a_*(\varphi,\psi,\iota+kh;h)u(\psi)\leqno(A.8)$$
where as usual, $a_*(\varphi,\psi,\iota;h)=a(\varphi,\iota;h)$ for (2,1)-quantization, 
$a_*(\varphi,\psi,\iota;h)=a(\psi,\iota;h)$ for (1,2)-quantization, or $a_*(\varphi,\psi,\iota;h)=a({1\over2}(\varphi+\psi),\iota;h)$
for Weyl quantization. We denote by $\widetilde L^m_{\delta}({\bf T}^d)$ the class of corresponding operators.
The generalisation of (A.8) to functions microlocally defined on
${\bf T}^d$, satisfying Floquet periodicity condition (1.6) will be considered in Sect.c).

To conclude this Section, we recall from [Roy] (and references therein), 
some properties of $h$-PDO's with periodic coefficients, which extend in a natural way
the calculus on ${\bf R}^d$. Note that we can replace Fourier series by Fourier transforms if we lift functions defined on ${\bf T}^d$
to ${\bf R}^d$ by using local exponential charts (see [CdV1]). 

\medskip
\noindent $\bullet$ {\it Fourier series}.
A symbol $a(\varphi,\iota)$ belongs to $\widetilde S^m_{\delta}({\bf T}^d)$ iff its Fourier series $\widehat a(k,\iota)$
satisfies the following estimate: For all $s\in{\bf N}$ and $\beta\in{\bf N}^d$, there is $C_{s,\beta}>0$ such that
$$|\partial_\iota^\beta a(k,\iota;h)|\leq 
C_{s,\beta}\langle k\rangle^{-s} h^{m-\delta|\beta|}, \quad k\in{\bf Z}^d$$
\smallskip
\noindent $\bullet$ {\it Asymptotic expansions}. 
Let $0\leq\delta<1$, and $\delta^* >0$, consider a sequence 
$a_j\in\widetilde S^{m+j\delta^* }_{\delta}({\bf T}^d)$, we say that $\Sum_ja_j$ is asymptotic to $a\in\widetilde S^m_{\delta}({\bf T}^d)$ and we note
as usual $a(\varphi,\iota;h)\sim \Sum_ja_j(\varphi,\iota;h)$ iff for each $J\in{\bf N}$, 
$$a(\varphi,\iota;h)-\Sum_{j=0}^{J-1}a_j(\varphi,\iota;h)\in \widetilde S^{m+J\delta^* }_{\delta}$$
When $\delta^*=1-\delta$, we say simply that $a$ is a $\delta$-classical symbol. Usual Borel resommation procedure ensures that 
each such $\Sum_ja_j$ admits an asymptotic sum $a$. 
\smallskip
\noindent $\bullet$ {\it Composition and commutators}. 
If $a,b\in\widetilde S^m_{\delta}({\bf T}^d)$, we define $a\sharp b$ (for (2,1)-quantization) by
$$(a\sharp b)(\varphi,\iota;h)=(2\pi)^{-d}\int_{{\bf T}^d}\, d\psi\Sum_{k\in{\bf Z}^d} e^{ik(\psi-\psi)}a(\varphi,\iota+kh;h)b(\psi,\varphi;h)
\leqno(A.10)$$
and the product $D=AB$ (Moyal product) has symbol $d(\varphi,\iota;h)=(a\sharp b)(\varphi,\iota;h)$. Moreover $d(\varphi,\iota;h)$
has the following $\delta$-classical expansion $a\sharp b\sim\Sum_{j=0}^\infty d_j$, where $d_j\in\widetilde S^{j(1-\delta)}_{\delta}({\bf T}^d)$
are given by: 
$$d_j(\varphi,\iota;h)=\Sum_{|\alpha|=j}{1\over\alpha!}(hD_\iota)^\alpha a(\varphi,\iota;h)\partial^\alpha_\varphi b(\varphi,\iota;h)\leqno(A.11)$$
Again, when $a(\varphi,\iota;h)$ is defined locally near $\iota=\iota_0$, we may replace the sum over $k$'s in (A.8)
by a finite sum. 

From this, we can easily obtain the symbol $c={i\over h}(a\sharp b-b\sharp a)$ of the commutator ${i\over h}[A,B]$, with principal term
equal to the Poisson bracket $\{a,b\}$. When $a\in\widetilde S^m_0({\bf T}^d)$ is independent of $\varphi$, 
and $b\in\widetilde S^m_{\delta}({\bf T}^d)$,
then $c$ has an expansion of the form:
$$c(\varphi,\iota;h)\sim\{a,b\}+\Sum_{j\geq2}\Sum_{|\alpha|=j}{i\over h}(hD_\iota )^\alpha
a(\iota;h)\partial^\alpha_\varphi b(\varphi,\iota;h)\leqno(A.12)$$
(see [Roy] for details). 
\smallskip
\noindent $\bullet$ {\it $L^2$ continuity and adjoints}. 
Every $a\in\widetilde S^0_{\delta}({\bf T}^d)$ gives a continuous operator $A$ on $L^2({\bf T}^d)$, its adjoint $A^*$
is a $h$-PDO in the same class, and its symbol, denoted by $a^*$ is given by
$$a^*(\varphi,\iota;h)=(2\pi)^{-d}\int_{{\bf T}^d}\, d\psi\Sum_{k\in{\bf Z}^d} e^{ik(\psi-\psi)}\overline{a(\psi,\iota+kh;h)}$$
\smallskip
\noindent $\bullet$ {\it Exponentials and adjoint representations}. For our purposes, we only need the case $\delta=0$. Let 
$P\in\widetilde L^m_0({\bf T}^d)$,and $B\in\widetilde L^0_0({\bf T}^d)$, then  
$C=e^{iP}Be^{-iP}\in\widetilde L^0_0({\bf T}^d)$, with $C\sim\Sum_{n\geq0}C_n$, $C_n\in\widetilde L^{(m+1)n}_0({\bf T}^d)$
is given by $C_n={i^n\over n!}[P,\cdots,[P,B]\cdots]$. 
\bigskip
\noindent{\bf c) Action-angle variables and quantization}
\smallskip
Here we address the problem of quantization in action-angle variables in a neighbhd of an invariant Diophantine torus.
Recall [BaWe, Definition 5.33] that a linear map $\rho$
from the space $C^\infty(T^*M)$ of smooth functions, to the algebra ${\cal A}$ 
generated by self-adjoint operators on some complex Hilbert space ${\cal H}$, and endowed with the Lie algebra structure defined by
$[A,B]_h={i\over h}(AB-BA)$, is called a {\it quantization} 
provided it satisfies so-called Dirac axioms~: (1) $\rho(1)=\Id$, (2) $\rho(\{f,g\})=[\rho(f),\rho(g)]_h$, (3) for some complete set of 
functions $f_1,\cdots,f_n$ in involution, the operators $\rho(f_1),\cdots,\rho(f_n)$ form a complete commuting set.
We know this set of axioms is in general too stringent, in the sense that a quantization of all classical observables doesn't exist,
although $h$-Fourier integral operators, provide sometimes a good framework for approximation of
this classical-quantum correspondence. 

Quantization deformation occurs already in the simple case of
a completely integrable Hamiltonian system on $T^*M$, with Hamiltonian $H(p,x)$, 
which admits a family of Lagrangian tori $\Lambda^I$. Namely, trying to quantize the corresponding 
action-angle variables $(I,\varphi)$, considered as classical observables, in a neighborhood of the $\Lambda^I$'s, we
require that $\rho(I)=\widehat I,\rho(\varphi)=\widehat\varphi$ would satisfy 
$$[\widehat I_j,\widehat I_k]=0, \quad [\widehat \varphi_j,\widehat I_k]_h=\delta_{jk}, 
\quad [\widehat \varphi_j,\widehat \varphi_k]_h=0\leqno(A.15)$$
and moreover, that the semi-classical Hamiltonian $\widehat H(x,hD_x)$ associated with $H(x,p)$ {\it via} usual 
Weyl $h$-quantization, would be a function of $(\widehat\varphi,\widehat I)$. The naive answer would consist in choosing 
$\widehat\varphi$ as multiplication by $\varphi$, and $\widehat I=hD_\varphi$. 
But then if we try to recover the canonical operators 
$\widehat x_j=X_j(\widehat\varphi,\widehat I),\widehat p_k=P_k(\widehat\varphi,\widehat I)$ by symbolic calculus,
it turns out that the canonical commutation relations 
$[\widehat p_k,\widehat x_j]_h=\delta_{jk}$ are only satisfied modulo ${\cal O}(h)$, as can be checked when $H$
is the harmonic oscillator, with $p=\sqrt I\cos\varphi,x=\sqrt I\sin\varphi, H(p,x)=p^2+x^2=I$. 
See however [CdVVu] for the case of commuting $h$-PDO's. Extending an argument of [CdV1], 
we reduce the problem to microlocal Floquet-Bloch theory.
\medskip
\noindent $\bullet$ {\it Microlocal Floquet-Bloch theory on the torus}.
\smallskip
Let ${\bf T}={\bf R}/2\pi{\bf Z}$, ${\bf T}^*={\bf R}/{\bf Z}$ (interpreted as a the first Brillouin zone)
and $\alpha\in{\bf Z}_4^d$ (interpreted as a set of Maslov indices). 
Consider first the direct decomposition 
$L^2({\bf R}^d)\approx\int^\oplus_{{\alpha\over4}+{\bf T}^{*d}}L^2_\theta({\bf T}^d)\, d\theta$, 
over the shifted torus ${\alpha\over4}+{\bf T}^{*d}$, and the map 
$$U:L^2({\bf R}^d;d\varphi)\to\int^\oplus_{{\alpha\over4}+{\bf T}^{*d}}L^2_\theta({\bf T}^d)\, d\theta, \quad w\mapsto(Uw)_\theta=v_\theta\leqno(A.20)$$
where $v_\theta$ satisfies Floquet periodicity condition
$$v_\theta(\varphi-2k\pi)=e^{2i\pi\langle\theta+{\alpha\over4},k\rangle}v_\theta(\varphi), \quad k\in{\bf Z}^d\leqno(A.21)$$
We have Fourier expansion
$$v_\theta(\varphi)=\Sum_{k\in{\bf Z}^d}e^{2i\pi\langle\theta+{\alpha\over4},k\rangle}w(\varphi+2k\pi)\leqno(A.22)$$ 
so that by Parseval identity
$\int_{{\alpha\over4}+{\bf T}^{*d}}d\theta\,\int_{{\bf T}^d}d\varphi\,|v_\theta(\varphi)|^2=\|w\|^2_{L^2({\bf R}^d)}$, showing 
easily that $U$ is unitary (see [ReSi,Vol.IV]). 

We introduce a semi-classical version $U^h$ of $U$, 
in replacing $v_\theta$ in (A.21) by $v^h_{\theta}$, summing now over the
$k\in{\bf Z}^d$ which verify $|k|h\leq C h^\delta$, $0<\delta<1$; the isometry and Floquet periodicity properties are broken,
but if $w\in{\cal S}({\bf R}^d)$, we have 
$$\leqalignno{
&\int_{{\alpha\over4}+{\bf T}^{*d}}d\theta\,\int_{{\bf T}^d}d\varphi\,|v^h_\theta(\varphi)|^2=
\|w\|^2_{L^2({\bf R}^d)}+{\cal O}(h^N)\|w\|^2_{L^2({\bf R}^d)}&(A.23)\cr
&v_\theta(\varphi-2k\pi)=e^{2i\pi\langle\theta+{\alpha\over4},k\rangle}v_\theta(\varphi)+{\cal O}(h^N)\|w\|_{L^2({\bf R}^d)}, 
\quad k\in{\bf Z}^d, \quad |k|h\leq C' h^\delta&(A.24)\cr
}$$
for any $N$, and provided $0<C'<C$. 
  
For each $\theta$, we consider the flat Hermitean line bundle $E(\theta)$ over ${\bf T}^d$ associated with the
real cohomology class
$\theta+\alpha/4\in H^1({\bf T}^d;{\bf R})$, 
and whose sections are identified with functions $v_\theta\in L^2_\theta({\bf T}^d)$
that satisfy (A.20); we define similarly $E_h(\theta)$ in the semi-classical version (A.24). Namely, it is convenient to rescale the action by
$\widetilde\theta=h\theta\in{\bf T}^{*d}$, then
an orthonormal basis of 
$F_h(\widetilde\theta)=L^2_\theta({\bf T}^d;E_h(\widetilde\theta))$ consists of sections 
$$e_k^h(\varphi;\widetilde\theta)=\exp[-i\langle kh+\widetilde\theta+\alpha h/4,\varphi\rangle/h], \quad |k|h\leq C h^\delta \leqno(A.25)$$
(Bloch functions). These sections lift to $w^h:{\bf R}^d\to {\bf C}$ in the sense of (A.23). 
Following App.A.a, let $w^h\in L^2({\bf R}^d;d\varphi)$, and $\rho\in T^*{\bf R}^d$, 
we say that $\rho\notin\WF ^\delta w^h$ iff there is an admissible box
$\Pi=\Pi_{\rho}^{\delta,\nu,\gamma}$ centered at $\rho$, with $\nu\geq c$, $\gamma\geq c h^\delta$ for some $c>0$,
such that $w^h$ is negligible in $\Pi$. We have:
\medskip
\noindent {\bf Lemma c.1}: {\it Let $Z$ be the zero-section of ${\bf T}^d\times{\bf T}^{*d}\subset T^*{\bf T}^d$. 

i) For any 
$C>0$, $\WF ^\delta e_k\subset Z$, uniformly in $k$, $|k|h\leq C h^\delta$. In other words, for all
$h^{\delta}$-neighbhd $\Omega_*^h$ of 0 in ${\bf T}^{*d}$
$$\bigcup_{\widetilde\theta\in\Omega_*^h}\WF ^\delta e_k^h(\cdot,\widetilde\theta)\subset Z,  \quad |k|h\leq C h^\delta $$

ii) For all 
$w^h\in\int^\oplus_{{\alpha h\over4}+\Omega_*^h}F_h(\widetilde\theta)\, {d\widetilde\theta\over h}$, we have $\WF ^\delta w^h\subset{\bf R}^d\times0$.

iii) Conversely, if $\WF ^\delta w^h\subset \bigcup_{\{k:|k|h\leq ch^\delta\}}(2\pi k+{\bf T}^d)\times0$,
then $w^h\in\int^\oplus_{{\alpha h\over4}+\Omega_*^h}F_h(\widetilde\theta)\, {d\widetilde\theta\over h}$.}
\smallskip
\noindent {\it Proof}: i) follows from the very definition (A.25). Let $v_\theta^h\in F_h(\widetilde\theta)$, 
$v_\theta^h(\varphi)=\Sum_{|k|h\leq C h^\delta}a_k\exp[-i\langle kh+\widetilde\theta+\alpha h/4,\varphi\rangle/h]$, 
(with $a_k=a_k(\theta)$ normalized in $\ell^2$),  
and $\chi\in C^\infty_0({\bf R}^d)$
equal to 1 in a (fixed) neighbhd of $\varphi_0\in{\bf R}^d$.
In an exponential chart, 
we have ${\cal F}_h(\chi v_\theta^h)(\iota)=\Sum_{|k|h\leq C h^\delta}a_k(\theta)\widehat\chi\bigl({1\over h}(\iota+kh+\widetilde\theta+\alpha h/4)\bigr)$.
So integrating over the shifted torus gives 
$${\cal F}_h\int(\chi v_\theta^h)(\iota){d\widetilde\theta\over h}=
\Sum_{|k|h\leq C h^\delta}\int {d\widetilde\theta\over h}\, a_k(\theta)\widehat\chi\bigl({1\over h}(\iota+kh+\widetilde\theta+\alpha h/4)\bigr)$$
Using the fact that $\widehat\chi$ is rapidly decreasing, we see that if $|\iota|\geq C_1h^\delta$ for $C_1>0$ large enough,
then ${\cal F}_h(\chi w^h)(\iota)={\cal F}_h\int(\chi v_\theta^h)(\iota){d\widetilde\theta\over h}={\cal O}(h^\infty)$. 
This proves ii). Finally iii) follows from Fourier inversion formula. $\clubsuit$.
\smallskip
We may instead consider semi-classical distributions microlocalized near a given section of ${\bf R}^d\times{\bf T}^{*d}$,
namely for $I\in{\bf T}^{*d}$, replace in (A.25) $\widetilde\theta$ by $I+\widetilde\theta$.
It it straightforard to extend Lemma c.1 to that case.
\medskip
\noindent $\bullet$ {\it Semi-classical states on a manifold}.
\smallskip
Let now $M$ be a $d$ dimensional smooth manifold, and $y=(x,p)$ be local symplectic coordinates on $T^*M$.  
For $J=(J_1,\ldots,J_d)\in\neigh(I^0;{\bf R}^d)$, let
$$i(J):{\bf T}^d\to T^*M, \quad \varphi\mapsto y=(X(\varphi,J), P(\varphi,J))\leqno(A.26)$$ 
be a smooth family of embeddings, 
such that $\Lambda(J)=i(J)({\bf T}^d)$ is a Lagrangian torus, parametrized by  
angle coordinates $\varphi$, which define the half-density $|d\varphi|^{1/2}$ on $\Lambda(J)$.    
Identifying ${\bf T}^d$ with its image through $i(J)$, we will denote again by $i(J):\Lambda(J)\to T^*M$ the corresponding Lagrangian embedding.
Let $(\gamma_j)_{1\leq j\leq d}$ be basic cycles on
$\Lambda(J)$. They determine action variables $I_j=\oint_{\gamma_j} pdx$, and also the
Maslov indices  $\alpha_j$. Actions $J$ and angles
$\varphi$ on $\Lambda(J)$ are conjugated symplectic coordinates. The cycles $\gamma_j=\gamma_j(J)$ depending smoothly on 
$J$, Maslov indices $\alpha=(\alpha_1(J),\cdots,\alpha_d(J))$ are constant. 

This holds in case of Darboux-Weinstein theorem, i.e. when there is  
canonical transformation
$$\widetilde\kappa:\neigh(\Lambda^0;T^*M)\to\neigh(\iota=0;T^*{\bf T}^d)\leqno(A.27)$$
which maps $\Lambda^0$ to the zero section in $T^*{\bf T}^d$; 
here we have $J=I^0+\iota$, and write also $\Lambda(\iota)$ for 
$\Lambda(J)$. As usual we denote $\Omega_{1/2}^{\Lambda(\iota)}$ 
the bundle of half-densities on $\Lambda(\iota)$, and also by ${\bf L}^{\Lambda(\iota)}$ 
Maslov bundle on $\Lambda(\iota)$, whose holonomy is represented by the reduction modulo ${\bf Z}$ of the real cohomology class
$$I^{\Lambda(\iota)}/h+\alpha/4\in H^1(\Lambda(\iota);{\bf R})\leqno(A.28)$$
$I^{\Lambda(\iota)}_j=\oint_{\gamma_j^{\Lambda(\iota)}} pdx$ being computed along fundamental cycles $\gamma_j^{\Lambda(\iota)}$ over $\Lambda(\iota)$.

Recall the space of semi-classical states, or Lagrangian distributions on a manifold $M$:
\medskip
\noindent {\bf Definition c.2}: {\it Let $M=M^d$ be a smooth manifold, ${\bf H}_M$ an Hermitean bundle over $M$, and $\Lambda\subset T^*M$ an 
embedded Lagrangian manifold, parametrized locally by a non degenerate phase function $S(x,\eta)$, $\eta\in{\bf R}^d$
i.e. 
$$\Lambda=\{(x,{\partial S\over\partial x}): {\partial S\over\partial \eta}=0\}$$
Let $a\in S^m(M;{\bf H}_M)$, we call Lagrangian distribution associated with $\Lambda$ an oscillatory integral of the form}
$$I_h(a,S)(x)=(2\pi h)^{-d/2}\int a(x,h)e^{iS(x,\eta)/h}\, d\eta\leqno(A.29)$$
The set of such Lagrangian distributions is denoted by ${\cal I}^m(\Lambda;{\bf H}_M)$, and we say that 
$\Lambda$ is parametrized locally by $I_h(a,S)$. The sections of the Hermitean bundle ${\bf H}_M$ consist first of the tensor product 
of half-densities
$\Omega_{1/2}^{\Lambda}$ over $\Lambda$ with sections ${\bf L}^{\Lambda}$ of Maslov line bundle.
Following [BaWe] we call it the {\it intrinsic Hilbert space} over $M$, since it contains also half-densities. 
Maslov line bundle has transition functions $\exp i\pi(\sgn S''-\sgn\widetilde S'')/4$
for a change of phase function $S$ in $\Lambda_S\cap\Lambda_{\widetilde S}$, 
and $\exp i\pi\bigl( \sgn S''_{(x,\theta),(x,\theta)}-\sgn$ $S''_{(x',\theta),(x',\theta)}\bigr)/4$
for a change of coordinates $x\mapsto x'$ in $\Lambda_S$.
With $I_h(a,S)$ we associate its ``principal oscillating symbol'' of the form 
$$e^{i\Phi_S(\xi)/h}A_0(\xi)=e^{i\Phi_S(\xi)/h}e^{i\pi\sgn S''/4}a_0\bigl(x(\xi),\theta(\xi)\bigr)
\sqrt{\delta_S}$$ 
by a partition of unity subordinated to the covering of $\Lambda$ by the local charts $\Lambda_S$.
Here $a_0$ is the principal symbol of $a$,
and $A_0(\xi)=e^{i\pi\sgn S''/4}a_0\bigl(x(\xi),\theta(\xi)\bigr)\sqrt{\delta_S}\in \Omega_{1/2}^{\Lambda}\otimes{\bf L}^{\Lambda}$.
In addition, the amplitude $I_h(a,S)$ may be valued in an Hermitean vector space $E_M$, for instance the span of $L^2$ functions of 
type (A.25) above, microlocalized on $\Lambda(\iota)$. So we take ${\bf H}_M=\Omega_{1/2}\otimes{\bf L}\otimes E_M$.

When $\Lambda$ is quantizable, i.e. Maslov class $I^{\Lambda(\iota)}/h+\alpha/4\in {\bf Z}^d$, then (A.28) defines a distribution globally
on $\Lambda$, but it needs not be so.

Considerations above apply to the families of tori $\Lambda(J',N)\approx{\bf T}^d\times\{J'=I+\iota'\}$ constructed while taking
the classical Hamiltonian $H$ to its BNF. 
For simplicity, we have assumed that $H$ has no sub-principal symbol $H_1$, in which case one should add to $I^{\Lambda(J',N)}$ the integral
over $\Lambda(J',N)$ of the sub-principal 1-form $\langle H_1\rangle_{J',N}{d\varphi\over2\pi}$, see [DoRo,Thm1.2].

We know [DoRo,Corollary 2.4] that the action integral over a fundamental cycle $\gamma_j$ of 
$\Lambda(J',N)$ satisfies
$$I^{\Lambda(J',N)}_j={1\over  2\pi}\oint_{\gamma_j} P(J',\varphi')\,dX(J',\varphi')={1\over  2\pi}\oint_{\gamma^0_j} P^0(\varphi)\,dX^0(\varphi)
+\iota'_j+{\cal O}(\iota'^2)\leqno(A.31)$$
so that $d\theta={1\over h}(1+{\cal O}(\iota'))d\iota'$, showing that the map $\iota'\mapsto\theta$ is a local isomorphism; 
thus $\theta$ ranges over a period in ${\bf T}^{*d}$ as $\iota'$ varies of order $h$, i.e. 
as we move from a quantizable torus $\Lambda(J',N)$ (i.e. $I^{\Lambda(J',N)}/h+\alpha/4\in {\bf Z}^d$)
to nearby ones, so when $|\iota'|\leq h^\delta$, $\theta$ covers about $h^{\delta-1}$
times the torus ${\bf T}^{*d}$ (or Brillouin zone). 
Again, we could associate Maslov canonical operator with each quantizable torus
$\Lambda(J',N)$
(see e.g. [Laz] or [DoRo] for a simpler proof); 
this suffices to provide a sequence of quasi-modes, but not a ``global'' reduction of $H$ 
(i.e. microlocally in a $h^{\delta/2}$-neighbhd of Diophantine torus $\Lambda$) to an operator acting only in the $\varphi$ variable. 

Through $i(J)$ the flat Hermitean bundles $E(\theta)$ (resp. $E_h(\theta)$,~)
over ${\bf T}^d$ identify with a 
flat Hermitean bundle over $\Lambda(\iota)$, still denoted by $E(\iota)$ (resp. $E_h(\iota)$), by setting $\theta=I^{\Lambda(\iota)}/h$.
In particular, for any such $\iota$, $H^1(\Lambda(\iota);{\bf R})$ identifies with $H^1\bigl(\widetilde\kappa(\Lambda(\iota));{\bf R}\bigr)$. 
\medskip
\noindent $\bullet$ {\it Composition of semi-classical states and microlocal Floquet-Bloch theory on a manifold}.
\smallskip
Let $X=X^d$ be another smooth manifold, following [BaWe] we call the map $S_{M,X}:\overline{T^*M}\times T^*X\to T^*(M\times X)$
defined in local coordinates by $\bigl((x,\xi),(\varphi,\iota)\bigr)\mapsto (x,\varphi,-\xi,\iota)$ the {\it Schwartz transform}.
Here $\overline{T^*M}$ is simply $T^*M$ endowed with $-\sigma_M$. 
We have $(S_{M,X})^*\sigma_{M\times X}=\sigma_M\oplus-\sigma_X$. 
So if $\widetilde\kappa:T^*M\to T^*X$ a canonical transformation as in (A.27), call
$C_{\widetilde\kappa}\subset T^*M\times T^*X$ its graph (canonical relation), then 
$L_{\widetilde\kappa}=\{(x,\varphi,\xi,-\iota): (x,\xi,\varphi,\iota)\in C_{\widetilde\kappa}\}$ 
is Lagrangian for the canonical
2-form $\sigma_M\oplus\sigma_X$, and parametrized locally by a non degenerate phase function $S(x,\varphi,\eta)$, $\eta\in{\bf R}^d$, i.e.
$$L_{\widetilde\kappa}=\{(x,{\partial S\over\partial x},\varphi,-{\partial S\over\partial \varphi}): {\partial S\over\partial \eta}=0\}\leqno(A.32)$$
Let $a\in S^m(X\times M;{\bf H}_X\otimes{\bf H}_M)$, we call Lagrangian distribution associated with $L_{\widetilde\kappa}$ 
an oscillatory integral of the form
$$I_h(a,S)(x,\varphi)=(2\pi h)^{-d/2}\int a(x,\varphi;h)e^{iS(x,\varphi,\eta)/h}\, d\eta\leqno(A.33)$$
The set of such Lagrangian distributions is denoted by ${\cal I}^m(X\times M;L_\kappa;{\bf H}_X\otimes{\bf H}_M)$.
If $\Lambda\subset T^*M$ is a Lagrangian manifold, then $C_{\widetilde\kappa}\circ\Lambda$ is Lagrangian in $T^*X$.
In particular, we say that the family 
$\Lambda=\Lambda(\iota)=i(\iota)({\bf T}^d)$ as in (A.26), foliating a neighbhd of some $\Lambda^0\subset T^*M$, is {\it parametrized} by
oscillating integrals (or Lagrangian distributions)
$I_h(a,S)(x,\varphi)$ on local charts $\Lambda_S$, which can be chosen independent of $\iota$ if $\iota$ is small enough.
The phase functions $S$ are such that $\iota={\partial S\over\partial \varphi}$ on the critical set ${\partial S\over\partial \eta}=0$.
Note that here $\Lambda$ and $C_\kappa\circ\Lambda$ are transverse in $T^*(X\times M)$, but 
composition of semi-classical states holds in more general situations called ``clean intersection'' (see [We], [BaWe,Sect.6] for details.~) 
To (A.33) corresponds
(modulo smoothing operators) an operator 
$$A:C_0^\infty(X;{\bf H}_X)\to C^\infty(M;{\bf H}_M)\leqno(A.34)$$
with Schwartz kernel $K_A(x,\varphi;h)=I_h(a,S)(x,\varphi)$.
In other words, $I_h(a,S)(x,\varphi)$ is the Schwartz kernel of an operator in $\Hom({\bf H}_X,{\bf H}_M)$. 

Note that in the discussion above, we can replace $\iota$ by $\iota'$ when having replaced $\widetilde\kappa$ by $\widetilde\kappa\circ\kappa_N$
at the $N$th step of the BNF. 
\medskip
\noindent $\bullet$ {\it Intertwining property}.
\smallskip
We adapt the constructions of [We] and [CdV1] (see also [MeSj]) to show first that for each $\theta\in{\bf T}^{*d}$, 
there is a partially isometric FIO $A_\theta$ intertwining the $h$-PDO $H(x,hD_x)$ on $L^2(M)$ with a $h$-PDO $P_\theta$ on 
$L^2_\theta(X)$, $X={\bf T}^d$ such that $\Ran P_\theta\subset F_h(\theta)$. 
This follows from the semi-classical BNF as in [HiSjVu,Sect.3], or [Roy], which we review here. 

Using (A.27) and (A.31) we can replace in the notations $\theta$ by $\iota'$;
we drop also sometimes the prime in variables $\varphi',\iota',J'$.
For definiteness, let us fix some terminology. We call the sequence of $h$-PDO's  with symbols $P_N(\varphi,\iota,h)\in S^0({\bf T}^d)$ 
{\it nested} microlocally near $\Lambda$, 
if for all $N$, $P_{N+1}(\varphi,\iota,h)-P_N(\varphi,\iota,h)={\cal O}\bigl(|\iota_1,h|^{N+1}\bigr)$. A nested sequence of $h$-PDO's also
admits an asymptotic sum $P$.

We call again the sequence of  FIO's $U_N$  {\it nested}, if for all $N$, $\|U_{N+1}-U_N\|={\cal O}(h^{N+1})$. 
If $P_N, G^{(N)}$ are nested, so is 
$$e^{i\ad\,G^{(N)}}P_N=e^{iG^{(N)}}P_Ne^{-iG^{(N)}}$$
Given a sequence of (unitary) nested FIO's $U_N$ we can always construct an asymptotic 
FIO $U$ such that for all $N$, $\|U-U_N\|={\cal O}(h^{N+1})$. We have:
\medskip
\noindent {\bf Proposition c.3} (BNF): {\it Let $\Lambda=\Lambda^0$ be a Lagrangian torus with Diophantine frequencies as above, 
$(\varphi,\iota)$ a system of action-angle coordinates defined in a neighborhood of $\Lambda$ as in (A.27), and $H(x,\xi;h)$ 
a classical symbol on $T^*M$ microlocally defined near $\Lambda^0$
(we denote the corresponding $h$-PDO by the same letter.)
Then there exists a nested sequence $P_N$ of $h$-PDO's  
defined microlocally near $\Lambda$, a nested sequence of elliptic FIO's $U_N$, 
such that the Weyl symbol
$P_N$ (denoted by the same letter) of $U_N^{-1}HU_N$ verifies
$$P_N(\varphi',\iota';h)=P_N(\iota',h)+{\cal O}(|\iota',h|^{N+1})\leqno(A.37)$$
More precisely, for each $N$, there exists $g^{(N)}(\varphi',\iota')=\Sum_{j=1}^N g_j(\varphi',\iota')$
homogeneous of degree $j+1$ in $\iota$, and $G^{(N)}(\varphi',\iota';h)=\Sum_{j=0}^N h^jG_j(\varphi',\iota')$, such that on the operator
level
$$e^{i\ad\,G^{(N)}}e^{\ad\,g^{(N)}/h}H=P_N(hD_{\varphi'},h)+R_{N+1}(\varphi',hD_{\varphi'},h)\leqno(A.38)$$
where the full symbol of $P_N(hD_{\varphi'},h)$ is independent of $\varphi'$ and $R_{N+1}(\varphi',\iota';h)={\cal O}(h^{j+1}+|\iota'|^{N+1})$.}
\smallskip
\noindent {\it Sketch of proof}:
We have constructed already a nested sequence of canonical transformations $\kappa_N$ near $\Lambda$
that take the classical Hamiltonian $H$ to its Birkhoff normal form near $\Lambda$, i.e. for each $N\geq 1$,
a smooth function $g_N(\varphi',\iota')$ generating a 
canonical transformation $\kappa_N=\exp X_{g_N}$, such that
$H\circ\widetilde\kappa\circ\kappa_N(\varphi',\iota')=H_N(\iota')+{\cal O}(\iota'^{N+1})$, and
$H_N(\iota')=H|_{\Lambda}+\langle\omega,\iota'\rangle+{\cal O}(\iota'^2)$.
When $N=1$, $(\varphi,\iota)=(\varphi',\iota')$. 
Call $S=S_N(x,\varphi',\eta)$ 
a generating function for $\widetilde\kappa\circ\kappa_N$ in a suitable chart, so that 
$\iota'={\partial S\over\partial \varphi'}$ on the critical set ${\partial S\over\partial \eta}=0$.
Taking $a_0=1$ in (A.33), we have already constructed a FIO $V_N=A_{\iota',N,0}$ (with $j=0$) such that, by Egorov theorem, 
the Weyl symbol of $V_N^{-1}HV_N$
takes the form $H_N(\iota')+{\cal O}(\iota'^{N+1})+{\cal O}(h)$. In other words, there are
new action-angle coordinates $(\varphi',\iota')$ relative to which the 
$h$-PDO's $P_N=e^{\ad\,S_N/h}H=V_N^{-1}HV_N$ has principal symbol $H_N(\iota')$ modulo ${\cal O}(|\iota'|^{N+1})$. 
Then we proceed to the higher order corrections in $h$ of the BNF, by conjugating $V_N$ by elliptic $h$-PDO's.
This is done as in [HiSjVu,Sect.3], solving homological equations along the flow of $\exp X_{g_N}$, 
modulo errors ${\cal O}(|\iota'|^{N+1})$ (see also [DoRo,Sect.1] for similar constructions). 
(The symbolic calculus of $h$-PDO's and FIO's on $T^*{\bf T}^d$ (having periodic coefficients in $\varphi$), is reviewed in App.A.b.
It is suitable to work in exponential charts where we can make use of Fourier transform instead of Fourier series, see [CdV1].~)
Thus we can find a nested sequence of $h$-PDO's $G^{(N)}=\Sum_{j=0}^{N}h^jG_j$ with periodic coefficients, 
defined microlocally in a nghbd of $\Lambda$, such that (A.38) holds with $U_N^h=V_Ne^{iG^{(N)}}$. $\clubsuit$
\medskip
When $\iota={\cal O}(h^\delta)$, $0<\delta<1$, we can thus arrange so that the remainder verifies
$R_{N+1}(\varphi',\iota';h)={\cal O}(h^{N+1})$. 

Consider now the fibre bundle $i(J')_*(F_h(\iota'))$ over $M$, where $i(J')=i(J',N):X\to T^*M$, $i(J')(X)=\Lambda(\iota',N)$ is the embedding (A.26),
and let ${\bf H}_M^{\iota'}=\Omega_{1/2}^{\Lambda(J')}\otimes{\bf L}^{\Lambda(J')}\otimes i(J')_*(F_h(\iota'))$.
The Schwartz kernel $K_{\iota',N}$ of $U_N^h$
has the form (A.33), with phase $S=S_N(x,\varphi',\eta)$, in particular $\iota'={\partial S_N\over\partial\varphi')}$
is constant on $\Lambda(J',N)$. So
$$K_{\iota',N}\in{\cal I}^0\bigl(X\times M;L_{\widetilde\kappa\circ\kappa_N};{\bf H}_X^\theta\otimes{\bf H}_M^{\iota'}\bigr)\leqno(A.40)$$
with $\theta=\theta(\iota')$. As in [CdV1,Lemme 10.3], we have the following intertwining property:
\medskip
\noindent {\bf Corollary c.4}: {\it Given $j\geq0$, for $N$ sufficiently large, we can find $\Omega^h$ a $h^\delta$-neighbhd of $\iota'=0$
(and a corresponding $h^\delta$-neighbhd $\Omega_*^h$ of $\widetilde\theta=0$,~)
a smooth family $\bigl(U_{\iota',N,j}\bigr)_{\iota'\in\Omega}$ of OIF with Schwartz kernel
$K_{\iota',N,j}\in{\cal I}^0$, partially isometric in $\Hom({\bf H}^\theta_X,{\bf H}_M^{\iota'})$,
and a smooth family $P_{N,j}(hD_\varphi;h)$ of self-adjoint $h$-PDO whose symbols depend on $\iota'$ only, such that
$$\leqalignno{
&H(x,hD_x)U_{\iota',N,j}-U_{\iota',N,j}P_{N,j}(hD_{\varphi'};h)\in{\cal I}^{-(j+1)}
\bigl(X\times M;L_{\widetilde\kappa\circ\kappa_N};{\bf H}_X^\theta\otimes{\bf H}_M^{\iota'}\bigr)&(A.41)\cr
&U_{\iota',N,j}^*U_{\iota',N,j}-\id\in{\cal I}^{-(j+1)}
\bigl(X\times X;\id;{\bf H}_X^\theta\otimes{\bf H}_X^\theta\bigr)&(A.42)\cr
}$$
uniformly for $(\theta,\iota')\in\Omega_*^h\times\Omega^h$.}
\smallskip
We take also asymptotic sums $U_{\iota'}$ (resp. $P$) for $A_{\iota',N,j}$ 
(resp. $P_{N,j}$) with respect to $N$ and $j$. 
By construction of the BNF (see the proof of Proposition c.3), $P_{\widetilde \theta}=P_{\iota'}=P(hD_{\varphi'};h)$
is independent (modulo ${\cal O}(h^\infty)$) of $\iota'$. 
Consider the fiberwise superposition of the corresponding semi-classical states $K_{\iota'}$
$$A=\int A_{\widetilde\theta}\, {d\widetilde\theta\over h}:\int^\oplus_{{\alpha h\over4}+\Omega_*^h}F_h(\widetilde\theta){d\widetilde\theta\over h}
\to L^2(M)\leqno(A.44)$$
so that, by (A.41) 
$$HA=H\int A_{\widetilde\theta}\, {d\widetilde\theta\over h}=\int A_{\widetilde\theta}P_{\widetilde \theta}\, {d\widetilde\theta\over h}=
\bigl(\int A_{\widetilde\theta}\, {d\widetilde\theta\over h})P(hD_{\varphi'};h)=AP(hD_{\varphi'};h)$$
modulo negligible operators in the sense of App.A.a (i.e. with ${\cal O}(h^\infty)$ kernels), if $\Omega_*^h$
is a sufficiently small $h^\delta$-neighbhd of $I^0=I(\Lambda^0)$. Using Lemma c.1 and composing with $U^h$ as in (A.20) we eventually proved:
Theorem 1.1. $\clubsuit$
\smallskip
Note that we recover the semi-classical spectrum of $H(x,hD_x)$ in a $h^\delta$-neighbhd of $I^0=I(\Lambda^0)$, using $A_{\widetilde\theta}$
when $I^{\Lambda(\iota')}/h+{\alpha/4}\in{\bf Z}^d$, in which case $A_{\widetilde\theta}$ is well defined (and univalued) on the torus
(see e.g. [HiSjVu]). 
When the sub-principal symbol $H_1$ is non zero, we have to slightly modify (A.28) and the construction above 
to account for the sub-principal 1-form.
\bigskip
\centerline {\bf References}
\medskip
\noindent [ArKoNe]  V.Arnold, V.Kozlov, A.Neishtadt. Mathematical aspects of classical and celestial mechanics. Encyclopaedia of Math. Sci.,
Dynamical Systems III, Springer, 2006.

\noindent [BaWe] S.Bates, A.Weinstein. Lectures on the geometry of quantization. Berkeley Math. Lect. Notes 88,
American Math. Soc. 1997.

\noindent [BeDoMa] V.Belov, S.Dobrokhotov, V.Maksimov. Explicit formulas for  generalized action-angle variables
in a neighborhood  of an isotropic torus and their applications. Theor. Math. Phys., 135(3),p.765-791, 2003.

\noindent [Bi] D.Birkhoff, Dynamical Systems. American Math. Soc. Colloquium Publ. Vol.IX, Providence, Rhode Island, Rev.ed, 1966.

\noindent [Bo] J.B. Bost. Tores invariants des syst\`emes dynamiques hamiltoniens. S\'eminaire Bourbaki, Expos\'e 639, Vol. 1984-85.

\noindent [BrDoNe] J.Br\"uning, S.Dobrokhotov, R.Nekrasov. Quantum dynamics in a thin film 2. Russian J. Math. Phys. 16(4), p.467-477, 2009.

\noindent [ChSh] S.S.Chern, Z.Shen. Riemann-Finsler Geometry. Nankai Tracts in Math. World Scientific, 2005.

\noindent [CdV]  Y.Colin de Verdi\`ere. {\bf 1}. Modes et quasi-modes sur les vari\'et\'es riemanniennes. Inventiones Math.
43, p.15-52, 1977. {\bf 2}. M\'ethode de moyennisation en M\'ecanique
semi-classsique. Journ\'ees Equations aux D\'eriv\'ees partielles, Expos\'e No 5, Saint Jean de Monts, 1996.

\noindent [DoRo] S.Dobrokhotov, M.Rouleux. {\bf 1}. 
The semi-classical Maupertuis-Jacobi correspondence for quasi-periodic hamiltonian flows with applications to linear 
water waves theory. Asymptotic Analysis, Vol.74 (1-2), p.33-73, 2011.
{\bf 2}. The semi-classical Jacobi-Maupertuis correspondence: stable and unstable spectra.
Proceedings ``Days of Diffraction 2012'', Saint-Petersburg. IEEE 10.1109/ DD.2012.6402752, p.59-64. 

\noindent [DoZh] S.Dobrokhotov, A.Shafarevich. ``Momentum'' tunneling between tori and the splitting of 
eigenvalues of the Laplace-Beltrami operator on Liouville surfaces. Math. Phys. Anal. Geometry 2, p.141-177, 1999. 

\noindent [Du] C.Duval. Spinoptics. Comm. Math. Phys. 

\noindent [EasMat] M.Eastwood, V.M.Matveev. Metric connexions in projective geometry. IMA Vol. Math. Appl., 144, Springer, 2008.

\noindent [FeGe] C.Fermanian-Kammerer, P.G\'erard. Mesures semi-classiques et croisement de modes. Bull. Soc. Math. France 130(1), p.123-168, 2002.

\noindent [GeLe] P.G\'erard, E.Leichtnam. Ergodic properties of eigenfunctions for the Dirichlet problem. Duke Math. J. 71(2), p.559-607, 1993.

\noindent [deGo] M.de Gosson. Symplectic Geometry and Quantum Mechanics. Birkh\"auser, 2006.
 
\noindent [HeSj] B. Helffer, J. Sj\"ostrand. Multiple wells in the
semi-classical limit I. Comm. Part. Diff. Eqn. 9(4) p.337-408, 1984.

\noindent [HeMaRo] B.Helffer, A.Martinez, D.Robert. Ergodicit\'e et limite semi-classique. Comm.Math.Phys. 109, p.313-326, 1987.

\noindent [HiSjVu] M.Hitrik, J.Sj\"ostrand, S.Vu-Ngoc. Diophantine tori and spectral asymptotics for non-self adjoint 
operators. Amer. J. Math.  129(1), p.105-182, 2007. 

\noindent [H\"o] L.H\"ormander. The Analysis of Linear Partial Differential Operators III. Springer, 1985.

\noindent [Iv] V.Ivrii.  Microlocal Analysis and Precise Spectral Asymptotics. Springer-Verlag, Berlin,  1998.

\noindent [KaMa] M.V.Karasev, V.P.Maslov. Pseudodifferential operators and canonical operator 
in general symplectic manifolds. Math. USSR Izvestjia 23, p.277-305, 1984.

\noindent [Laz] V.Lazutkin. KAM theory and semi-classical approximation to eigenfunctions. Springer, 1993.

\noindent [Ma] V.P.Maslov. The complex WKB ethod for nonlinear equations I: Linear Theory. Birkh\"auser, Basel, 1994.

\noindent [MatTo] V.S.Matveev, A.Topalov. Geodesic equivalence via integrability. Geom. Dedicata, 96, p.91-115, 2003.

\noindent [MeSj] A.Melin, J.Sj\"ostrand. Bohr-Sommerfeld quantization condition for non self-adjoint operators in dimension 2.
Ast\'erisque 284, p.181-244, 2003.

\noindent [O-de-Al] A.Ozorio de Almeida. Hamiltonian Systems: Chaos and Quantization. Cambridge Univ. Press, 1990.

\noindent [PeR] V.Petkov, D.Robert. Asymptotique semi-classique du spectre d'Hamiltoniens quantiques et trajectoires
classiques p\'eriodiques. Comm.Part.Diff.Eqn's 10(4), p.365-390, 1985.

\noindent [Ral] J.V.Ralston. On the construction of quasi-modes associated with periodic orbits. Comm. Math. Phys. 51(3) p.219-242, 1976.

\noindent [Roy] N.Roy. A semi-classical KAM theorem. Comm.Part.Diff.Eqn's 32(5), p.745-770, 2007.

\noindent [Rui] S.Ruijsenaars. The Aharonov-Bohm effect and scattering theory. Ann. Phys. 146, p.1-34, 1983.

\noindent [Sh] A.V.Shnirelman. On the asymptotic proporties of the eigenfunctions in the region of chaotic motion.
Addendum to [Laz].

\noindent [Ta] M.Taylor. Finsler structures and wave propagation, {\it in}: V.Isakov (ed.), Sobolev spaces in Mathematics III.
Int. Math. Series, Springer, 2009.

\noindent [We] A.Weinstein. On Maslov quantization condition, {\it in}:  Fourier Integral Operators and 
Partial Differential Equations. J.Chazarain, ed. Lecture Notes in Math. 469, Springer, p.361-372, 1974. 

\noindent [Zi] W.Ziller. Geometry of the Katok examples. Ergod. Theor. Dyn. Sys. 3, p.135-157, 1982.

\bye